\pdfoutput=1
\documentclass[twocolumn]{aastex63}
\usepackage[T1]{fontenc}
\usepackage[utf8]{inputenc}

\usepackage{amsmath,amsthm,amssymb}
\usepackage{graphicx}
\usepackage{lipsum}
\usepackage{hyperref}
\usepackage{xcolor}
\usepackage{bm}
\usepackage{textcomp}
\usepackage{needspace}
\let\tablenum\relax

\usepackage{siunitx}
\usepackage{natbib}
\nonstopmode

\newcommand{\kms}{km\,s$^{-1}$}

\newcommand{\degree}{$^{\circ}$}
\newcommand{\um}{\textmu m}

\newcommand{\cothreetwo}{$^{12}$CO(J=3$\rightarrow$2)}
\newcommand{\coonezero}{$^{12}$CO(J=1$\rightarrow$0)}
\newcommand{\nurest}{$\nu_{\mathrm{rest}}$}
\newcommand{\inv}{$^{-1}$}
\newcommand{\hst}{\textit{HST}}
\newcommand{\gaia}{\textit{Gaia}}
\newcommand{\Var}{\mathrm{Var}}

\newcommand{\uat}[2]{\href{http://astrothesaurus.org/uat/#2}{#1 (#2)}}
\DeclareUnicodeCharacter{2212}{-}

\begin{document}

\title{Molecular gas in a gravitationally lensed galaxy group at $\bm{z=2.9}$}

\author[0000-0001-6662-7306]{Jeff Shen}
\affiliation{Dunlap Institute for Astronomy and Astrophysics, University of Toronto, 50 St George Street, Toronto ON, M5S 3H4, Canada}
\affiliation{Department of Statistical Sciences, University of Toronto, 100 St George Street, Toronto ON, M5S 3G3, Canada}

\author[0000-0003-2475-124X]{Allison W. S. Man}
\affiliation{Dunlap Institute for Astronomy and Astrophysics, University of Toronto, 50 St George Street, Toronto ON, M5S 3H4, Canada}
\affiliation{Department of Physics \& Astronomy, University of British Columbia, 6224 Agricultural Road, Vancouver, BC V6T 1Z1, Canada}

\author[0000-0002-9842-6354]{Johannes Zabl}
\affiliation{Univ Lyon, Univ Lyon1, Ens de Lyon, CNRS, Centre de Recherche Astrophysique de Lyon UMR5574, F-69230, Saint-Genis-Laval, France}
\affiliation{Institute for Computational Astrophysics and Department of Astronomy \& Physics, Saint Mary’s University, 923 Robie Street, Halifax, Nova Scotia, B3H 3C3, Canada}

\author[0000-0002-7299-2876]{Zhi-Yu Zhang}
\affiliation{School of Astronomy and Space Science, Nanjing University, Nanjing 210093, People’s Republic of China}
\affiliation{Key Laboratory of Modern Astronomy and Astrophysics (Nanjing University), Ministry of Education, Nanjing 210093, People’s Republic of China}

\author[0000-0001-5983-6273]{Mikkel Stockmann}
\affiliation{Cosmic Dawn Center (DAWN)}
\affiliation{Niels Bohr Institute, University of Copenhagen, Jagtvej 128, DK-2100 Copenhagen, Denmark}

\author[0000-0003-2680-005X]{Gabriel Brammer}
\affiliation{Cosmic Dawn Center (DAWN)}
\affiliation{Niels Bohr Institute, University of Copenhagen, Jagtvej 128, DK-2100 Copenhagen, Denmark}

\author[0000-0001-7160-3632]{Katherine E. Whitaker}
\affiliation{Cosmic Dawn Center (DAWN)}
\affiliation{Department of Astronomy, University of Massachusetts, 710 North Pleasant Street, Amherst, MA 01003, USA}

\author[0000-0001-5492-1049]{Johan Richard}
\affiliation{Univ Lyon, Univ Lyon1, Ens de Lyon, CNRS, Centre de Recherche Astrophysique de Lyon UMR5574, F-69230, Saint-Genis-Laval, France}

\correspondingauthor{Jeff Shen}
\email{jshenschool@gmail.com}

\begin{abstract}
  Most molecular gas studies of $z>2.5$ galaxies are of intrinsically bright objects, despite the galaxy population being primarily "normal" galaxies with less extreme star formation rates. \deleted{Obtaining }Observations of normal galaxies at high redshift \deleted{will }provide a more representative view of galaxy evolution and star formation\added{, but such observations are challenging to obtain}. In this work, we present ALMA \cothreetwo\ observations of a sub-millimeter selected galaxy group at $z=2.9$, resulting in spectroscopic confirmation of seven images from four member galaxies. These galaxies are strongly lensed by the MS 0451.6–0305 foreground cluster \added{at z=0.55}, allowing us to probe the molecular gas content \deleted{in a regime that is otherwise time-costly to study}\added{on levels of $10^{9}-10^{10}~{\rm M_\odot}$}. Four detected galaxies have molecular gas masses of $(0.2-13.1)\times 10^{10}~{\rm M_{\odot}}$, and the non-detected galaxies have inferred molecular gas masses of $<8.0\times 10^{10}~{\rm M_{\odot}}$. \deleted{We calculate the molecular gas depletion time and investigate its redshift evolution using a compiled sample of 546 galaxies up to $z=5.3$.}\added{We compare these new data to a compilation of 546 galaxies up to $z=5.3$, and find that depletion times decrease with increasing redshift.} \added{We then compare the depletion times of galaxies in overdense environments to the field scaling relation from the literature, }\deleted{We find that for galaxies in overdense environments, higher redshift galaxies have shorter depletion times than those at lower redshifts. We investigate the impact of the environment on depletion time, but do not find statistically significant relations due the lack of CO detections of protocluster galaxies at $z>2.5$ with normal star formation rates.}\added{and find that the depletion time evolution is steeper for galaxies in overdense environments than for those in the field.} More \deleted{observations of this population}\added{molecular gas measurements of normal galaxies in overdense environments at higher redshifts ($z>2.5$) }are needed \deleted{in order }to verify the environmental dependence of star formation and gas depletion.

\end{abstract}


\received{2020 October 27}
\revised{2021 May 17}
\accepted{2021 May 20}

\section{Introduction}\label{sec:introduction} 

Molecular gas is the fuel for star formation and is consequently a key parameter in determining how galaxies evolve. Although we have a deep understanding of the galaxy populations of today \citep[see review by][]{Blanton_2009}, observations of distant galaxies are important for us to infer how galaxies formed and evolved \citep{Tacconi2020}. A crucial cosmic epoch is the peak of cosmic star formation history from $z\approx1-3$ when roughly half of the stars in the present-day Universe formed \citep{Shapley_2011,Madau_2014}.

High-redshift galaxies have historically been difficult to detect, relative to local galaxies, due to their distance. As such, \deleted{the conspicuous nature of the starburst population means a disproportionately large number of studies of molecular gas in high-redshift galaxies are of intrinsically very bright sources}\added{detections of molecular gas in high-z star-forming galaxies are heavily biased towards the most extremely star-forming galaxies, which harbour copious amounts of molecular gas} \citep[e.g.,][]{Walter_2004, Carilli_2010, Danielson_2010, Riechers_2010a, Combes_2012, Spilker_2015, Miller2018, Tadaki2018, Ciesla_2020, Spingola_2020}.

The vast majority of galaxies are not as strongly star forming as submillimeter galaxies \citep[SMGs;][]{Madau_2014}. It is thus vital to characterize the molecular gas of distant "normal" galaxies in order to understand the representative mode of star formation. Significant progress toward this goal has been made since the advent of the Atacama Large Millimeter/submillimeter Array (ALMA), whose sensitivity has enabled studies that probe faint carbon monoxide emissions (CO, a common tracer for molecular gas) in distant galaxies \citep[e.g., review by][]{Hodge2020}. However, obtaining CO detections of normal galaxies at $z>2$ remains a challenge.

To date, only a few galaxy clusters with molecular gas mass measurements have spectroscopic confirmations, with the majority of them at $z<2.5$ \citep{Noble2017, Rudnick_2017, Hayashi2018}. These cluster galaxies also present a range of results on the gas content, with gas fractions ($f_{gas} = \mathrm{M_{gas}} / (\mathrm{M_{gas} + M_{\ast}})$), where $\mathrm{M_{gas}}$ is the gas mass and $\mathrm{M_{\ast}}$ is the stellar mass) and depletion times ($t_{\mathrm{depl}} = \mathrm{M_{gas}} / \mathrm{SFR}$, where $\mathrm{SFR}$ is the star formation rate) varying by over an order of magnitude. In contrast, protoclusters discovered at $z>4$ all contain massive starburst galaxies totalling star formation rates (SFRs) of over $6000~{\rm M_\odot\,yr^{-1}}$ in the entire protocluster \citep{Miller2018, Oteo2018}, which may suggest that the $z=2-4$ is a time of transition in the star formation of galaxy protoclusters.

A relevant and ongoing discussion relating to molecular gas in high-redshift galaxies is the environmental dependence of molecular gas properties like the gas fraction and depletion time. Generally, a high-density environment means that a higher molecular gas content, and thus star formation, would be expected \added{\citep{Schmidt_1959}}. However, the dense environment also promotes more interactions between galaxies and facilitates processes such as ram-pressure stripping and strangulation which would reduce a galaxy's gas content during their infall towards the cluster center \added{\citep{Mo_2010}}. Some studies find evidence of higher molecular gas fractions in clusters than field scaling relations (\citealp{Noble2017, Hayashi2018, Gomez-Guijarro_2019}; \citealp[][for galaxies with stellar mass $10.5 < \log{(\mathrm{M_\ast} / \mathrm{M_\odot})} < 11.0$]{Tadaki2019}). On the other hand, there is also evidence for similar molecular gas fractions between coeval cluster and field galaxies \citep{Dannerbauer_2017, Lee_2017, Stach_2017}, and for the presence of massive galaxies, primarily near the center of clusters, which contain extremely low molecular gas fractions \citep{Wang2018, Zavala2019}. This molecular gas deficiency is attributed to quenching mechanisms which inhibit gas accretion or cooling; alternatively, the molecular gas may have been rapidly consumed or expelled \citep[][and references therein]{Man_2018}. However, most of these studies on the molecular gas content of cluster galaxies have been limited to $z<2.5$ \citep{Tacconi2020}. Obtaining higher redshift observations of protocluster galaxies would shed light on the impact of an overdense environment on molecular gas, and how that evolves with time.

The use of strong gravitational lensing makes it possible to detect galaxies that are otherwise too faint for spectroscopic follow up. Previous studies have found success using lensing as a tool to probe \deleted{normal }galaxies \deleted{which have}\added{with} relatively low\deleted{er} SFRs ($\ll 1000~{\rm M_\odot\,yr^{-1}}$)\added{, which are considered to be "normal" at $z>2$} \citep{Saintonge_2013, Dessauges_Zavadsky_2015, Dessauges_Zavadsky_2017}.

In this paper, we present \cothreetwo\ observations of a group of interacting normal star-forming galaxies at $z\approx2.9$ in a giant sub-millimeter arc \citep{Borys_2004} lensed by the MS 0451.6--0305 foreground cluster \citep{MacKenzie2014}. The group is composed of nine sources, \deleted{labelled}\added{labeled} Gals. 1-9 as in \cite{MacKenzie2014}. We note that Gal. 9 is a foreground galaxy, and do not present an analysis of its CO content in this work. The total SFR, as derived from far-infrared (FIR) luminosity, of $450\pm50~{\rm M_\odot\,yr^{-1}}$ for the entire group is far lower than the SFR of known protoclusters at comparable redshifts; the protoclusters in \cite{Oteo2018}, \cite{Miller2018}, \cite{Wang2018}, \cite{Gomez-Guijarro_2019}, and \cite{Zavala2019} have SFR that range from $1400~{\rm M_\odot\,yr^{-1}}$ to over $6000~{\rm M_\odot\,yr^{-1}}$. Several of the background \deleted{submm}\added{sub-millimeter (submm)} sources in the group are multiply imaged due to the lensing, and have been observed at optical and radio wavelengths \citep{Borys_2004, Berciano_Alba_2007, Berciano_Alba_2010, Zitrin_2010, Jauzac2020}.

The environment, redshift, and SFR of the lensed sources make them valuable for investigating star formation in early protocluster galaxies. Using data from ALMA, we examine the molecular gas content of the sources and compare our results to a compilation of high-redshift galaxies. Using this compiled sample, we also discuss the impact of the environment and the redshift evolution of the depletion time, an important quantity in the discussion of molecular gas underpinning the future star formation potential of galaxies.

The layout of this paper is as follows. In Section \ref{sec:observations} we describe the observations and data reduction. In Section \ref{sec:results} we analyze the CO observations and describe the calculation of molecular gas masses of the sources. In Section \ref{sec:discussion} we discuss our findings in the context of high-redshift galaxy evolution, compare the gas properties of our sources against a compilation of primarily CO-detected high-redshift galaxies, and consider the uncertainties in our results. A summary of this paper is presented in Section \ref{sec:conclusion}.

Throughout this paper, we assume a \cite{Kroupa_2001} initial mass function (IMF), a flat $\Lambda$CDM cosmology, a Hubble constant at $z=0$ of $H_0 = 70~{\rm km\,s^{-1}\,Mpc^{-1}}$, and a density of non-relativistic matter at $z=0$ of $\Omega_M=0.3$, in units of the critical density. \added{At ${z=2.9}$, 1\arcsec\ corresponds to $7.78~{\rm kpc}$.}

\Needspace{4\baselineskip}
\section{Observations}\label{sec:observations} 

\Needspace{4\baselineskip}
\subsection{ALMA}\label{sec:alma}

Observations of MS 0451.6--0305 were performed with the ALMA 12\,m array under the project code 2017.1.00616.S (PI: Allison Man). The 41 antennas were arranged in a compact configuration with baselines ranging from 14\,m to 783\,m. The total on-source integration time was 64.08 minutes over two consecutive execution blocks on 12--13 March 2018. The precipitable water vapor during the observations was about 2.94--3.19\,mm. J0423-0120 was used as a flux and bandpass calibrator while J0501-0159 was used for phase calibration. The pointings were centered on Gal. 1.b following the \citet{MacKenzie2014} naming convention. We used ALMA Band 3 in a mixed spectral setup to target the \cothreetwo\ transition (\nurest = 345.796\,GHz) at $z\approx2.9$, the redshift of the lensed group. A spectral window was centered on $88.136~{\rm GHz}$ covering a bandwidth of $1.875~{\rm GHz}$ with 240 channels. The native channel width is thus 26.6\,\kms.

The data were calibrated with the Common Astronomy Software Package \citep[CASA version 5.4.0;][]{McMullin2007} with pipeline version 42030.
The data cube was imaged with natural weighting to optimize the signal-to-noise (SNR) of compact sources.
The resulting synthesized beam is 1.81\arcsec\,$\times$\,1.28\arcsec\, corresponding to 14.08\,kpc\,$\times$\,9.96\,kpc at $z=2.9$, with a position angle of $-64.3$\degree.
We further bin the cube with the \texttt{CLEAN} task \citep{Hogbom_1974} to a channel width of 100\,\kms\ to improve SNR.
\deleted{The root mean square (rms) noise is estimated from a 3\arcsec\-radius source-free circular aperture near the pointing center of the primary-beam-corrected data cube.
The resulting rms noise is  0.117\,mJy\,beam\inv\ for the 100\,\kms-resolution cube and 0.241\,mJy\,beam\inv\ for the native-resolution cube. }The moment maps in Figure \ref{fig:hst} are created with the non-primary-beam corrected cube, and the spectra are extracted from the primary-beam corrected cube. We measure different noises for each source, as the sources are spread out over the field-of-view.

\Needspace{4\baselineskip}
\subsection{\hst}\label{sec:hst}

\added{\textit{Hubble Space Telescope} (\hst) wide-field imaging observations of MS 0451.6--0305 are available in six optical or near-infrared filters \citep{Egami2010,Jauzac2020}. The main purposes of using \hst\ images for this work is to create an RGB image (Figure \ref{fig:hst}) and to identify the source positions for spectral extraction (Section \ref{sec:results-specextraction}). For these purposes, we use the deepest optical image (F814W) taken with the Advanced Camera for Surveys (ACS), and the two near-infrared images (F110W and F160W) taken with the Wide-Field Camera 3 (WFC3). Three epochs of ACS/F814W images are available (program IDs: 9836, 10493, 11591) for a total exposure time of 11438\,s. The WFC3 F110 and F160W images each has exposure time of 2612\,s and 2412\,s, respectively (program ID 11591). The individual exposures were retrieved from the Mikulski Archive for Space Telescopes website\footnote{\url{https://archive.stsci.edu}}. For each ``visit'' (i.e., a set of exposures of a target taken with an instrument and filter combination at each epoch), the absolute astrometry is aligned to stars in the \gaia\ DR2 catalog \citep{GAIADR2}. The alignment of the \hst\ images is done with seven \gaia\ DR2 stars within the \hst\ footprint, six of which are matched to better than 20\,mas. We have also compared the positional offsets of stars in the Pan-STARRS catalog PS 1 \citep{Chambers2019}, which has a higher source density but lower astrometric precision than \gaia. For the 55 stars from Pan-STARRS PS1, the residual offset is 43\,mas. Both of these values are smaller than the 60\,mas size of one \hst\ pixel. Standard imaging calibration procedure were applied with the \texttt{grizli} analysis software \citep{Brammer2019}\footnote{\url{https://github.com/gbrammer/grizli}}. The exposures per filter are combined into a mosaic and drizzled onto a common pixel scale of 0.06\arcsec.}

\Needspace{4\baselineskip}
\section{Analysis and Results}\label{sec:results} 

\Needspace{4\baselineskip}
\subsection{Star Formation Rates}\label{sec:results-sfr}

The SFRs used in this paper are obtained from \cite{MacKenzie2014}, who employed a forward-modelling approach in combination with Markov Chain Monte Carlo methods to \added{simultaneously deblend and }fit spectral energy distributions (SEDs) to the source galaxies whose emissions are blended together in a giant submillimeter arc. They made use of 450\,\um\ and 850\,\um\ data from the Submillimetre Common-User Bolometer Array (SCUBA)-2, as well as \textit{Herschel} Spectral and Photometric Imaging Receiver (SPIRE) and Photodetecting Array Camera and Spectrometer (PACS) data. \added{The locations of \hst\ candidate counterparts are used as priors for the source of submm emissions during the SED fitting. }By integrating the resulting SEDs, they obtain intrinsic FIR luminosities, which were converted to SFRs \added{with the relation
  \begin{align}
    \mathrm{SFR} = 1.49\times 10^{-10}~{\rm M_\odot\,yr^{-1}\,L_{FIR}\,L_\odot^{-1}}
  \end{align}
from \cite{Murphy2011} with the assumption that UV radiation is completely absorbed by dust and reradiated at longer wavelengths}. We refer the reader to \cite{MacKenzie2014} for a more detailed description of the procedures.

\Needspace{4\baselineskip}
\subsection{Spectrum Extraction}\label{sec:results-specextraction} 

\begin{figure*}[!ht]
  \centering
  \includegraphics[width=\linewidth]{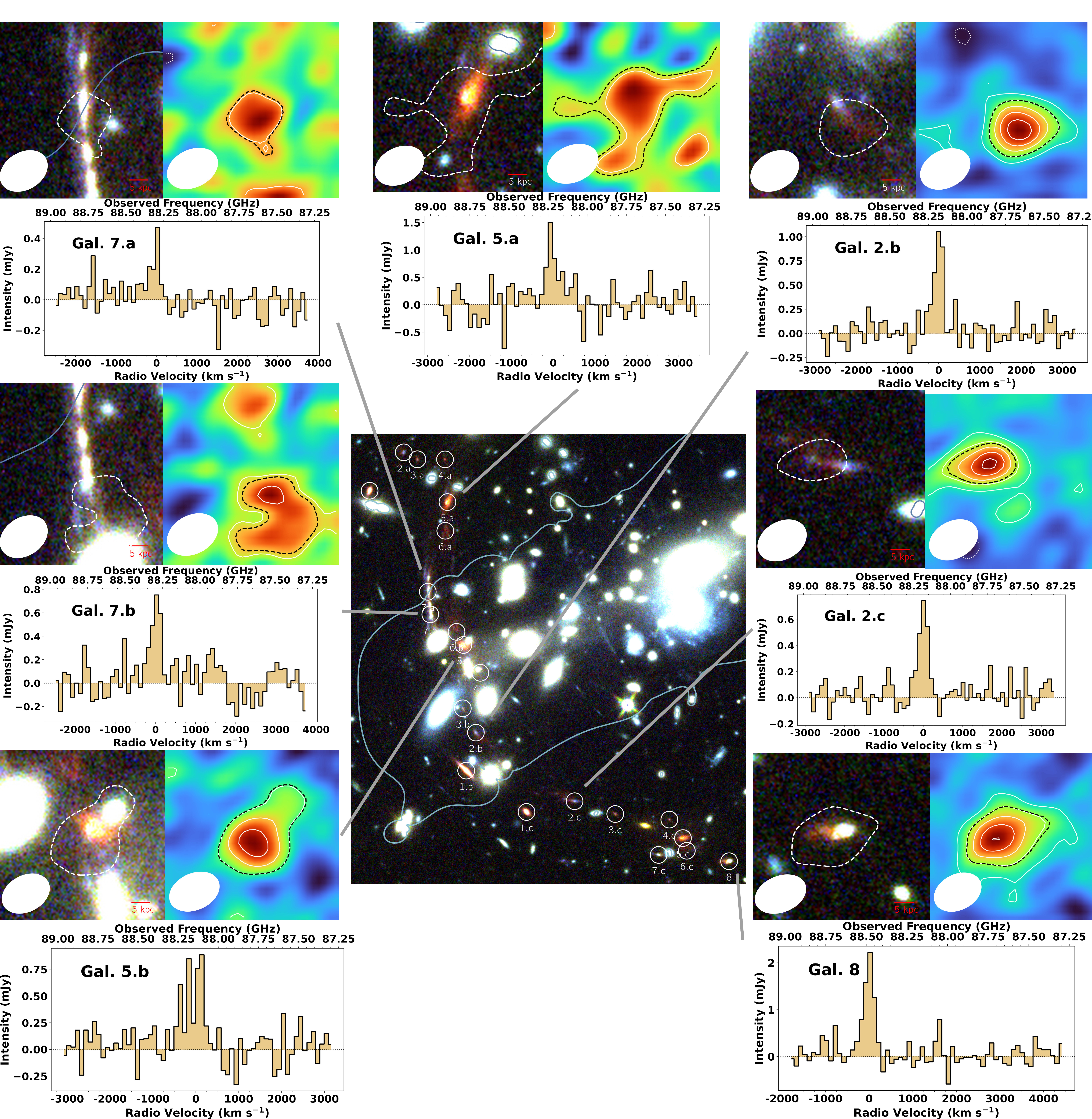}
  \caption{The central figure is a composite color image created using \hst\ F160W, F110W, and F814W filters as the red, green, and blue channels, respectively. \added{Overplotted in blue is the critical line of the lens model.}
    The moment zero maps are created from the $-200~{\rm km\,s^{-1}}$ to $200~{\rm km\,s^{-1}}$ channels of the non-primary-beam corrected cubes. The optimized extraction regions are indicated in the corresponding moment zero maps by the dotted black line. The solid white contours start from $2\,\sigma$ and go up in $2\,\sigma$ increments, and the dotted contours start from $-2\,\sigma$ and go in $2\,\sigma$ increments, where $\sigma$ is the rms noise. The ellipses in the bottom-left corners are the ALMA beam, which has FWHM major and minor axis sizes of 1.810\arcsec\ and 1.275\arcsec\ respectively, and a position angle of $-64.3~{\rm deg}$.
    The spectra are extracted by spatially integrating the primary-beam corrected cube within the optimized extraction regions in all \deleted{velocity channels}\added{channels in the spectral window}.
    Also included are cutouts of the \hst\ image with the contour corresponding to the CO extraction aperture overlaid in white.
  Each cutout measures 6\arcsec\,$\times$\,6\arcsec, corresponding to roughly $46~{\rm kpc} \times 46~{\rm kpc}$ \added{on the image plane} at $z=2.9$. In each cutout, the up direction is north and the left direction is east.}
  \label{fig:hst}
\end{figure*}

High-redshift galaxies can have slight offsets between their optical/near-infrared (IR) and submillimeter positions due to different astrometry reference systems or intrinsic offsets between stars and gas. In our data, we note offsets in the locations of the \hst\ peaks and the corresponding ALMA peaks. We explore reasons for the observed offsets in Section~\ref{sec:discussion-offset} and argue that the offsets are intrinsic, rather than \added{astrometric.}\deleted{systematic. The offsets are visible in the \hst\ cutouts in Figure \ref{fig:hst}.}

We \deleted{first }recalculate\deleted{d} the \hst\ \deleted{peaks}\added{locations} by \deleted{taking the centroid of 2D Gaussians that were fitted to the}\added{first creating an} \hst\ RGB image\deleted{, created} with \added{the} F160W, F110W, and F814W filters as the red, green, and blue channels, respectively. \added{We then use a combination of cropping and brightness masking to isolate each source as best as possible before fitting 2D Gaussians to each source using a least squares algorithm \citep{Levenberg1944} and taking the center as the \hst\ location. }These new locations are given in Table \ref{table:hstloc}. \deleted{They}\added{The \hst\ images} are aligned to the \deleted{Gaia}\added{\gaia} astrometric data and differ from the positions given in \cite{MacKenzie2014} by an average of \deleted{0.343}\added{0.344}\arcsec. This difference is due to the different astrometry reference systems as well as the filters used for locating the sources.

For each detected galaxy, the ALMA peak was calculated \deleted{by taking the pixel within the optimal extraction region (see below for a description of how this is obtained) where the CO emission is the strongest in the moment zero map.}\added{by applying the same 2D Gaussian fitting technique used to find the \hst\ locations to the moment zero maps (see below for a description of how these are obtained). The centers of the best-fit Gaussians are also reported in Table \ref{table:hstloc}.} The offset between the \hst\ and ALMA peaks for detected galaxies, where the CO emission is detected and both \deleted{peak }locations are unambiguous, are given in Table \ref{table:offset}. \deleted{Similar star-gas offsets have been observed by \cite{Carilli_2010}, \cite{Hodge_2013}, and \cite{Spingola_2020}.}

\deleted{The offsets are significant; at 0.6\arcsec\,$-$\,1.0\arcsec\ (equivalent to $4.6-7.8~{\rm kpc}$ at $z=2.9$), they are $3-5$ times larger than the typical astrometric precision of the \hst\ data.}

\added{For each galaxy with reported offsets, both the \hst\ and ALMA positions are passed through the \cite{MacKenzie2014} lens model along with their uncertainties to obtain corresponding positions and uncertainties on the source plane. From these source plane positions, we are able to calculate the offset and position angle on the source plane. These offsets also reported in Table \ref{table:offset}, range from 0.14\arcsec\ to 0.33\arcsec\, and are statistically significant for all five sources.}

\begin{deluxetable*}{ccccccccc}[ht!]
  \tablecaption{\hst\ and ALMA locations of sources. \label{table:hstloc}}
  \tablehead{
    \colhead{Gal} & \colhead{R.A.} & \colhead{R.A. Err.} & \colhead{Dec.} & \colhead{Dec. Err.} & \colhead{R.A.} & \colhead{R.A. Err.\tablenotemark{a}} & \colhead{Dec.} & \colhead{Dec. Err.\tablenotemark{a}} \\
    \colhead{ID} & \colhead{J2000} & \colhead{arcsec}  & \colhead{J2000} & \colhead{arcsec} & \colhead{J2000} & \colhead{arcsec} & \colhead{J2000} & \colhead{arcsec}
  }
  \startdata
  \multicolumn{1}{c}{} & \multicolumn{4}{c|}{\hst} & \multicolumn{4}{c}{ALMA\tablenotemark{e}} \\
  \hline
  1.a                  & 04$^{h}$ 54$^{m}$ 13.415$^{s}$ & 0.014\arcsec & -3$^\circ$\,00\arcmin\,42.708\arcsec & \multicolumn{1}{c|}{0.028\arcsec} & \nodata                        & \nodata      & \nodata                               & \nodata \\
  1.b                  & 04 54 12.643                   & 0.035        & -03 01 16.210                        & \multicolumn{1}{c|}{0.034}        & \nodata                        & \nodata      & \nodata                               & \nodata \\
  1.c                  & 04 54 12.161                   & 0.016        & -03 01 21.161                        & \multicolumn{1}{c|}{0.016}        & \nodata                        & \nodata      & \nodata                               & \nodata \\
  2.a                  & 04 54 13.142                   & 0.009        & -03 00 38.114                        & \multicolumn{1}{c|}{0.012}        & \nodata                        & \nodata      & \nodata                               & \nodata \\
  2.b                  & 04 54 12.565                   & 0.011        & -03 01 11.661                        & \multicolumn{1}{c|}{0.011}        & 04$^{h}$ 54$^{m}$ 12.537$^{s}$ & 0.181\arcsec & -03$^\circ$\,01\arcmin\,12.508\arcsec & 0.181\arcsec \\
  2.c                  & 04 54 11.777                   & 0.042        & -03 01 19.872                        & \multicolumn{1}{c|}{0.020}        & 04 54 11.841                   & 0.182        & -03 01 19.688                         & 0.181 \\
  3.a                  & 04 54 13.033                   & 0.009        & -03 00 38.943                        & \multicolumn{1}{c|}{0.015}        & \nodata                        & \nodata      & \nodata                               & \nodata \\
  3.b                  & 04 54 12.670                   & 0.026        & -03 01 08.755                        & \multicolumn{1}{c|}{0.037}        & \nodata                        & \nodata      & \nodata                               & \nodata \\
  3.c                  & 04 54 11.452                   & 0.016        & -03 01 21.404                        & \multicolumn{1}{c|}{0.016}        & \nodata                        & \nodata      & \nodata                               & \nodata \\
  4.a                  & 04 54 12.812                   & 0.009        & -03 00 38.957                        & \multicolumn{1}{c|}{0.011}        & \nodata                        & \nodata      & \nodata                               & \nodata \\
  4.b\tablenotemark{b} & \nodata                        & \nodata      & \nodata                              & \multicolumn{1}{c|}{\nodata}      & \nodata                        & \nodata      & \nodata                               & \nodata \\
  4.c                  & 04 54 11.020                   & 0.014        & -03 01 22.074                        & \multicolumn{1}{c|}{0.015}        & \nodata                        & \nodata      & \nodata                               & \nodata \\
  5.a                  & 04 54 12.793                   & 0.018        & -03 00 44.066                        & \multicolumn{1}{c|}{0.024}        & 04 54 12.807                   & 0.195        & -03 00 44.083                         & 0.199 \\
  5.b                  & 04 54 12.665                   & 0.061        & -03 01 01.168                        & \multicolumn{1}{c|}{0.044}        & 04 54 12.690                   & 0.183        & -03 01 01.726                         & 0.183 \\
  5.c                  & 04 54 10.911                   & 0.040        & -03 01 24.290                        & \multicolumn{1}{c|}{0.019}        & \nodata                        & \nodata      & \nodata                               & \nodata \\
  6.a\tablenotemark{c} & \nodata                        & \nodata      & \nodata                              & \multicolumn{1}{c|}{\nodata}      & \nodata                        & \nodata      & \nodata                               & \nodata \\
  6.b\tablenotemark{c} & \nodata                        & \nodata      & \nodata                              & \multicolumn{1}{c|}{\nodata}      & \nodata                        & \nodata      & \nodata                               & \nodata \\
  6.c\tablenotemark{c} & \nodata                        & \nodata      & \nodata                              & \multicolumn{1}{c|}{\nodata}      & \nodata                        & \nodata      & \nodata                               & \nodata \\
  7.a\tablenotemark{d} & \nodata                        & \nodata      & \nodata                              & \multicolumn{1}{c|}{\nodata}      & 04 54 12.934                   & 0.183        & −03 00 55.158                         & 0.182 \\
  7.b\tablenotemark{d} & \nodata                        & \nodata      & \nodata                              & \multicolumn{1}{c|}{\nodata}      & \nodata                        & \nodata      & \nodata                               & \nodata \\
  7.c                  & 04 54 11.105                   & 0.012        & -03 01 26.296                        & \multicolumn{1}{c|}{0.006}        & \nodata                        & \nodata      & \nodata                               & \nodata \\
  8                    & 04 54 10.542                   & 0.029        & -03 01 27.061                        & \multicolumn{1}{c|}{0.019}        & 04 54 10.577                   & 0.182        & -03 01 27.319                         & 0.181 \\
  \enddata
  \tablenotetext{a}{Calculated with Equations 7-9 from \cite{Schinnerer2010}.}
  \tablenotetext{b}{The source is obscured by a foreground cluster galaxy \citep{MacKenzie2014}.}
  \tablenotetext{c}{The diffuse structure of the source makes it difficult to determine an \hst\ position.}
  \tablenotetext{d}{The galaxy is resolved into multiple substructures in the \hst\ image.}
  \tablenotetext{e}{Positions are only provided for detected galaxies. A position for Gal. 7.b is not provided as its shape is highly irregular and somewhat diffuse.}
  \tablecomments{The positions of the sources \added{on the \hst\ image }are determined by using a least-squares fitting algorithm to fit a 2D Gaussian to \hst\ cutouts where the sources are isolated as best as possible. The uncertainties are the errors on the estimated means (i.e., not the standard deviations) of the fitted Gaussians. The \hst\ images are RGB composites created using the F160W, F110W, and F814W filters as the red, green, and blue channels, respectively. \added{ALMA source positions are determined by applying the same method to the $-200~{\rm km\,s^{-1}}$ to $200~{\rm km\,s^{-1}}$ moment-zero maps of the detected sources.}\deleted{The images are aligned to Gaia astrometry.}}
\end{deluxetable*}

\begin{deluxetable}{ccccccc}
  \tablecaption{Star-gas offset for detected galaxies. \label{table:offset}}
  \tablehead{
    \colhead{Gal} & \colhead{Offset} & \colhead{Position Angle} & \colhead{Offset} & \colhead{Position Angle}  \\
    \colhead{ID} & \colhead{arcsec} & \colhead{deg, east of north} & \colhead{arcsec} & \colhead{deg, east of north}
  }
  \startdata
  \multicolumn{3}{c|}{Image plane} & \multicolumn{2}{c}{Source plane} \\
  \hline
  2.b & $0.96_{-0.17}^{+0.18}$ & $207_{-11}^{+10}$ & $0.25_{-0.10}^{+0.11}$ & $98_{-7}^{+11}$ \\
  2.c & $0.99_{-0.18}^{+0.18}$ & $78_{-10}^{+12}$  & $0.25_{-0.13}^{+0.12}$ & $93_{-10}^{+11}$ \\
  5.a & $0.30_{-0.14}^{+0.17}$ & $104_{-50}^{+74}$ & $0.14_{-0.07}^{+0.11}$ & $95_{-27}^{+147}$ \\
  5.b & $0.70_{-0.18}^{+0.20}$ & $146_{-16}^{+15}$ & $0.25_{-0.08}^{+0.07}$ & $80_{-16}^{+14}$ \\
  8   & $0.61_{-0.17}^{+0.18}$ & $117_{-18}^{+18}$ & $0.33_{-0.10}^{+0.14}$ & $128_{-20}^{+27}$ \\
  \enddata
  \tablecomments{The offsets and position angles are calculated from the \hst\ locations to the ALMA locations (as given in Table \ref{table:hstloc}). \added{The estimates are obtained by perturbing the positions of each \hst\ and ALMA pair with Gaussian noise with standard deviation equal to the position uncertainties 1000 times, calculating the offset and position angle each time, and then taking the median of the sampling distribution. The uncertainties on these estimates are difference between the median and the 16 and 84 percentile values from the sampling distributions.}}
\end{deluxetable}

Because of the \hst--ALMA offsets, we do not extract spectra using apertures centered on the \hst\ positions. Rather, we adopt the following procedure to identify the optimal extraction apertures for the \cothreetwo\ spectra of each galaxy. We note that the non-detected galaxies have a different extraction procedure described later in the subsection.

For each source, we use an initial 6\arcsec\,$\times$\,6\arcsec\ extraction to create a velocity-integrated flux map (also known as moment zero map) centered at the \hst\ position. This is done by integrating the velocity channels ranging from $-200$\,\kms\ to $+200$\,\kms\ around the peak of the \cothreetwo\ emission as identified from the 1D spectrum, which is calculated by spatially integrating the same extraction. Contours of the local rms noise are also calculated and overplotted. This noise is calculated by fitting a 1D Gaussian to a histogram of a mirrored version of the negative flux values in a 60\arcsec\,$\times$60\arcsec\ moment zero map centered at the source. Examples of moment zero maps and contours are shown in Figure \ref{fig:hst}\added{, as well as in Appendix \ref{sec:appendix-m0}}.

The contours are used to optimize the extraction region from which we obtain the spectra. We use these contours to perform a masking procedure, which we repeat at $0.2\,\sigma$ increments from $0\,\sigma$ to $5\,\sigma$, where $\sigma$ is the rms noise as calculated previously. First, the primary contour corresponding to a given noise level (e.g., $2\,\sigma$) is selected and used to indicate the spatial extent of the CO emission in the original cube. Note that there may be multiple disjoint contours at this noise level; only the \textit{primary} contour which encloses the \deleted{CO emission peak}\added{pixel with the maximum integrated flux} is selected. The spectrum is then calculated by integrating the flux over the region within the contour in all \deleted{velocity channels}\added{channels in the spectral window}. \added{The SNR of this spectrum is defined as the flux in the peak channel over the rms flux of the spectrum with the peak channel removed.}

\deleted{Next, the SNR is calculated for this particular spectrum. The signal is defined as the average of the flux in the peak channel and the fluxes of the two adjacent channels, which ensures that the signal is coherent along the frequency axis and not a noise spike. We obtain the noise of the spectrum by removing the same three channels and calculating the rms flux from the rest of the channels within the spectral window. SNR is then the unitless ratio between the signal and the noise.}

Through this iterative process, we define an optimal extraction region (i.e., contour level) which maximizes the SNR of the spectrum. In Figure \ref{fig:hst}, the optimal extraction region of each source is given by the dashed black line in the moment zero maps, and by the (same) white line in the \hst\ cutouts. We note that this procedure is robust with respect to the initial extraction---regardless of the initial cube position or size (given that the source is entirely contained within the cube), the optimal region will be extracted.

Each source was inspected to ensure that there were no false signals due to contributions from nearby sources. Sources with $\mathrm{SNR > 3}$ as obtained by the described method were classified as a detection. The spectra of the detections, alongside the moment zero maps and \hst\ composite thumbnails, are shown in Figure \ref{fig:hst}.\added{ Integrated spectra for all galaxies can be found in Appendix \ref{sec:appendix-spec}.} The velocities of each spectrum are relative to the redshift of the respective galaxy as given in Table \ref{table:gasprops}. The displayed spectra are the ones with the highest SNR over the different extraction regions for each source.

For the non-detected galaxies with $\mathrm{SNR < 3}$ as calculated using the above procedure, we instead use a 1.5\arcsec-radius circular aperture centered at the \hst\ location to extract a subcube. The sources were assumed to be unresolved, and the spectra were calculated at the pixel position that had the maximum flux density on the $-200~{\rm km\,s^{-1}}$ to $200~{\rm km\,s^{-1}}$ moment zero map.

For each of the galaxies (both detections and non-detections), we also checked to make sure that the contribution from the continuum was negligible. The $-200~{\rm km\,s^{-1}}$ to $200~{\rm km\,s^{-1}}$ channels of each spectrum were masked out, and the mean and rms noise of the resulting line-free spectrum were calculated. In each case, the mean was consistent with zero (i.e., the absolute value of the mean was less than the rms noise), indicating that the continuum emission is insignificant. We also create a continuum image with the remaining three spectral windows where no \added{spectral line }emission is expected. None of the sources were detected in this continuum at $\mathrm{SNR > 5}$. The mean continuum noise at the location of the sources is 0.018\,mJy, and an upper limit of $3\,\sigma = 0.054~{\rm mJy}$ would be less than 10\% of the maximum flux density in each spectra (i.e., the flux density in the channel with the peak of the line).\added{ Continuum images are shown in Appendix \ref{sec:appendix-cont}.}

\Needspace{4\baselineskip}
\subsection{Molecular Gas Mass}\label{sec:results-gasmass} 

For the detected galaxies with $\mathrm{SNR > 3}$, each spectrum was fit with a 1D Gaussian model in order to obtain the \cothreetwo\ line flux, which was then used to calculate the molecular gas mass. In order to quantify the uncertainty in the measurement, a Monte Carlo simulation was performed. Each spectrum was perturbed with Gaussian noise, centered at the observed intensity and with a standard deviation equal to the rms noise of the signal. We then fit a Gaussian function to each of the 500 simulated spectra to obtain a distribution of best-fitting parameters.

The full width at half maximum (FWHM) for a Gaussian is related to its standard deviation $\sigma$ by $\text{FWHM} = 2\sqrt{2\ln{2}}\sigma$, and the line flux $S_{\mathrm{CO}(3\rightarrow2)}\Delta v$ is given by the area under the Gaussian curve. This is calculated analytically as $\sqrt{2\pi}a|\sigma|$, where $a$ is the peak intensity of the curve. The reported values and uncertainties for the peak frequency, FWHM line width, and flux in Table \ref{table:gasprops} are the means and standard deviations, respectively, of the Monte Carlo realizations.

For galaxies without a strongly detected \cothreetwo\ emission, we derive upper limits on the line flux by assuming a boxcar line profile. The width of the profile was taken to be 400\,\kms, which is slightly low relative to the median detection profile width of 512\,\kms. The height of the profile was set at $3\,\sigma$, where $\sigma$ is the rms noise of the 1D spectrum. The resulting $3\,\sigma$ upper limits on the line fluxes for the non-detections are provided in Table~\ref{table:gasprops}.

We measure spectroscopic redshifts from the \cothreetwo, $z_{\mathrm{CO}}$, which are more reliable than---but largely consistent with---the redshifts provided in \cite{MacKenzie2014} based on lensing predictions. Knowing that the rest frequency of the \cothreetwo\ transition is \nurest\,$\approx$\,345.796\,GHz \citep{Carilli2013}, we apply the equation $\nu_{\mathrm{obs}}=\nu_{\mathrm{rest}}/(1+z)$.

The \added{intrinsic} line luminosity \added{$L^\prime_{\mathrm{CO}}$} is calculated using the following equation from \cite{Solomon1992}\added{, which has been modified to include the lensing magnification}:
\begin{align}\label{eqn:linelum}
  L'_{\mathrm{CO}} = \frac{c^2}{2k} \frac{S_{\mathrm{CO}}\Delta v}{\added{\mu}} \frac{D_L^2}{\nu_{\mathrm{obs}}^2 (1+z)^3},
\end{align}
where \deleted{$L^\prime_{\mathrm{CO}}$ is the line luminosity for the \cothreetwo\ transition, }$k = 1.381\times 10^{-23}~{\rm J\,K^{-1}}$ is the Boltzmann constant, \added{$\mu$ is the magnification of the object (with $\mu=1$ for unlensed sources and $\mu>1$ for lensed sources), }$D_L$ is the luminosity distance in Mpc, and $\nu_{\mathrm{obs}}$ is the observed frequency of the emission in GHz.

This \cothreetwo\ luminosity was converted to an equivalent $^{12}$CO(J=1$\rightarrow$0) luminosity with an assumed rotational transition ratio of $L'_{\mathrm{CO}(3\rightarrow2)}/L'_{\mathrm{CO}(1\rightarrow0)}=0.5$ \citep{Tacconi2013}. The $^{12}$CO(1$\rightarrow$0) luminosity was then used to derive the molecular gas mass. We adopt a conversion factor of $\alpha_{\mathrm{CO}}=4.36~{\rm M_\odot / (K\,km\,s^{-1}\,pc^2})$, which accounts for the mass contribution from helium, following convention for Milky Way-like galaxies \citep{Bolatto_2013}. See Section \ref{sec:discussion-uncertainties} for further discussion of the selection and uncertainty of the spectral line energy distribution (SLED) and the $\alpha_{\mathrm{CO}}$ conversion factor. All of the resulting values are presented in Table \ref{table:gasprops}. \added{Note that the luminosity as defined above gives the intrinsic luminosity, whereas the values reported in Table \ref{table:gasprops} are observed (i.e., lensed) luminosities.}

\begin{deluxetable*}{chCCCCCCCCCC}[htb!]
  \tabletypesize{\scriptsize}
  \tablecaption{Molecular gas properties of galaxy images. \label{table:gasprops}}
  \tablehead{
    \colhead{Gal} & \nocolhead{Peak Frequency} & \colhead{FWHM} & \colhead{$S_{\mathrm{CO}(3\rightarrow2)}\Delta v$} & \colhead{RMS} & \colhead{$z$\tablenotemark{a}} & \colhead{$L'_{\mathrm{CO}(3\rightarrow2)}$} & \colhead{Amplification\tablenotemark{d}} & \colhead{Delensed $\rm M_{\mathrm{gas}}$} & \colhead{Delensed SFR\tablenotemark{d}} & \colhead{$t_{\mathrm{depl}}$}
    \\
    \colhead{ID} & \nocolhead{GHz} & \colhead{${\rm km\,s^{-1}}$} & \colhead{${\rm Jy\,km\,s^{-1}}$} & \colhead{$\times 10^{-4}~{\rm Jy\,beam^{-1}}$} & \colhead{} & \colhead{$\times 10^{10}~{\rm K\,km\,s^{-1}\,pc^{2}}$} & \colhead{} & \colhead{$\times 10^{10}~{\rm M_\odot}$} & \colhead{$\mathrm{M_\odot\,yr^{-1}}$} & \colhead{Gyr}
  }
  \startdata
  \multicolumn{11}{c}{Detections} \\
  \hline
  2.b & 88.19\pm0.01 & 260\pm49   & 0.305\pm0.038 & 1.3 & 2.9211\pm0.0005 & 1.3\pm0.2 & 8.1\pm0.4   & 1.4\pm0.2  & 99\pm9   & 0.14\pm0.02 \\
  2.c & 88.17\pm0.07 & 650_{-650}^{+1131} & 0.248\pm0.071 & 1.0 & 2.922\pm0.003 & 1.0\pm0.3 & 6.1\pm0.1   & 1.5\pm0.4  & 99\pm9   & 0.15\pm0.04 \\
  5.a & 88.24\pm0.02 & 446\pm264  & 0.503\pm0.150 & 3.2 & 2.919\pm0.001 & 2.1\pm0.6 & 5.3\pm0.1   & 3.4\pm1.0  & <35      & <0.98 \\
  5.b & 88.16\pm0.05 & 712_{-712}^{+992}  & 0.438\pm0.088 & 2.0 & 2.923\pm0.002 & 1.8\pm0.4 & 6.4\pm0.1   & 2.5\pm0.5  & <35      & <0.71 \\
  7.a & 88.43\pm0.14 & 982\pm890  & 0.155\pm0.064 & 1.0 & 2.910\pm0.006 & 0.6\pm0.3 & 33.0\pm2.0  & 0.2\pm0.1  & 11\pm2   & 0.15\pm0.07 \\
  7.b & 88.30\pm0.03 & 513\pm487  & 0.278\pm0.090 & 1.6 & 2.916\pm0.001 & 1.2\pm0.4 & 45.0\pm3.0  & 0.2\pm0.1  & 11\pm2   & 0.20\pm0.08 \\
  8   & 88.52\pm0.00 & 273\pm34   & 0.633\pm0.069 & 2.6 & 2.9065\pm0.0002 & 2.6\pm0.3 & 1.73\pm0.04 & 13.1\pm1.5 & 290\pm40 & 0.45\pm0.08 \\
  \hline
  \multicolumn{11}{c}{Non-Detections} \\
  \hline
  1.a & \nodata & \nodata & <0.424 & 3.6 & 2.92233_{-0.00010}^{+0.00013}\tablenotemark{b} & <1.7 & 3.8\pm0.06  & <3.9 & 4\pm1\tablenotemark{e} & \nodata \\
  1.b & \nodata & \nodata & <0.195 & 1.6 & 2.92233_{-0.00010}^{+0.00013}\tablenotemark{b} & <0.8 & 20.0\pm1.0  & <0.3 & 4\pm1\tablenotemark{e} & \nodata \\
  1.c & \nodata & \nodata & <0.195 & 1.6 & 2.92233_{-0.00010}^{+0.00013}\tablenotemark{b} & <0.8 & 7.3\pm0.1   & <1.0 & 4\pm1\tablenotemark{e} & \nodata \\
  2.a & \nodata & \nodata & <0.634 & 5.3 & 2.91\pm0.04 & <2.7 & 2.86\pm0.04 & <8.0 & 99\pm9             & <0.81 \\
  3.a & \nodata & \nodata & <0.475 & 4.0 & 2.94\pm0.04 & <2.0 & 3.19\pm0.05 & <5.4 & <23                & \nodata \\
  3.b & \nodata & \nodata & <0.222 & 1.9 & 2.94\pm0.04 & <0.9 & 2.98\pm0.05 & <2.7 & <23                & \nodata \\
  3.c & \nodata & \nodata & <0.259 & 2.2 & 2.94\pm0.04 & <1.1 & 4.31\pm0.08 & <2.2 & <23                & \nodata \\
  4.a & \nodata & \nodata & <0.364 & 3.0 & 2.94\pm0.04 & <1.5 & 3.57\pm0.06 & <3.7 & <50                & \nodata \\
  4.b & \nodata & \nodata & <0.231 & 1.9 & 2.94\pm0.04 & <1.0 & 6.2\pm0.2   & <1.3 & <50                & \nodata \\
  4.c & \nodata & \nodata & <0.299 & 2.5 & 2.94\pm0.04 & <1.3 & 3.36\pm0.06 & <3.2 & <50                & \nodata \\
  5.c & \nodata & \nodata & <0.334 & 2.8 & 2.89\pm0.03 & <1.4 & 2.89\pm0.04 & <4.2 & <35                & \nodata \\
  6.a & \nodata & \nodata & <0.364 & 3.0 & 2.86\pm0.03 & <1.5 & 8.2\pm0.2   & <1.6 & 53\pm14            & <0.30 \\
  6.b & \nodata & \nodata & <0.267 & 2.2 & 2.86\pm0.03 & <1.1 & 4.98\pm0.08 & <1.9 & 53\pm14            & <0.36 \\
  6.c & \nodata & \nodata & <0.323 & 2.7 & 2.86\pm0.03 & <1.4 & 2.76\pm0.04 & <4.2 & 53\pm14            & <0.80 \\
  7.c & \nodata & \nodata & <0.338 & 2.8 & 2.911\pm0.003\tablenotemark{c} & <1.4 & 2.87\pm0.04 & <4.2 & 11\pm2             & <3.86 \\
  \enddata
  \tablenotetext{a}{Redshifts for the detections are calculated from the CO detections. Redshifts for the non-detections are redshifts derived from the lensing model of \cite{MacKenzie2014} unless otherwise specified.}
  \tablenotetext{b}{Median $z_{spec}$ derived from VLT/X-SHOOTER spectrum of \added{Gal. }1.b (Man et al. submitted). The quoted lower and upper uncertainties are the differences between the median and the 16 and 84 percentiles, respectively.}
  \tablenotetext{c}{This is a spectroscopic redshift from \cite{Borys_2004}, derived from interstellar absorption lines.}
  \tablenotetext{d}{From \cite{MacKenzie2014} unless otherwise specified.}
  \tablenotetext{e}{Median SFR of Gal. 1.b from Man et al. submitted, adjusted to magnification $\mu=20$ from \cite{MacKenzie2014}. The quoted lower and upper uncertainties are the differences between the median and the 16 and 84 percentiles, respectively.}
  \tablecomments{Table of values for the detections (top portion of the table) and non-detections (bottom portion). The delensed gas mass\added{es and star formation rates are lensing-corrected values}\deleted{ is the intrinsic gas mass of the galaxy}. The rest of the data are observed values. Unless otherwise specified, values given with uncertainties are the means and standard deviations from Monte Carlo simulations. Upper limits are $3\sigma$ upper limits.}
\end{deluxetable*}

Since the sources are gravitationally lensed, we need to take into account the amplification factor for each image due to the lensing in order to calculate the delensed (intrinsic) molecular gas mass of each source. Although the amplification factor is given with respect to the flux density, all of the calculations necessary to go from flux density to the gas mass are linear in the relevant variables. Thus, we can directly apply the amplification factor to calculate the delensed gas mass.

We are aware of the revised lens model for the foreground cluster presented in \cite{Jauzac2020}, which would impact both the SFR and the molecular gas mass. For consistency in the magnification factors used for $\mathrm{M_{gas}}$ and SFR and to ensure that the depletion times are not affected by magnification, we choose to use the original magnifications as given in \cite{MacKenzie2014}. We note that although both $\mathrm{M_{gas}}$ and SFR are based on luminosities ($L^\prime_{\mathrm{CO}}$ and $L_{\mathrm{IR}}$ respectively), the scaling due to the change in magnification may not be equal between the two. This is due to the offsets between the CO emission and the SFR tracers, as described in Section \ref{sec:results-specextraction}. The different regions may be magnified differently by the foreground cluster, and thus one luminosity may be magnified more strongly than the other. We do not consider the effects of differential lensing in our analysis.

\Needspace{4\baselineskip}
\subsection{Spectrum Stacking}\label{sec:results-stack} 

Each galaxy in the group, apart from Gal. 8, has three multiple images. We stack the spectra from the three multiple images in an attempt to increase the SNR. Each of the spectra are normalized by their respective magnifications \citep{MacKenzie2014} so that the more strongly lensed sources do not inadvertently contribute more to the stacked result. In order to reduce the importance of noisier spectra, the stacking was done with an inverse-variance weighting (where the variances are also corrected for magnification), according to Equation \ref{eqn:stack}:

\begin{align}\label{eqn:stack}
  \mathbf{S_s} = \frac{1}{\sum_{i=1}^{3}\frac{1}{\Var(\mathbf{S_i})}} \sum_{i=1}^{3} \frac{1}{\Var(\mathbf{S_i})} \mathbf{S_i},
\end{align}
where the index $i = 1,2,3$ corresponds to the "a", "b", and "c" multiple images respectively,
$\mathbf{S_i}$ is the $\mathrm{i^{th}}$ delensed spectrum (i.e., $\mathbf{S_i} = \mathbf{s_i} / \mu_i$ where $\mathbf{s_i}$ is the $\mathrm{i^{th}}$ observed spectrum), $\Var(\mathbf{S_i})$ is the variance of the $\mathrm{i^{th}}$ delensed spectrum, and $\mathbf{S_s}$ is the stacked spectrum. The first term in the equation is a normalization term to ensure that the weights sum to 1.

For Gals. 2, 5, and 7, the stacking procedure increased the SNR over the maximum SNR of the constituent spectra (e.g., max SNR among Gals. 2.a, 2.b, and 2.c) by only a marginal amount. Furthermore, stacking the non-detected galaxies where none of the constituent sources were detected did not bring them over the $\mathrm{SNR > 3}$ threshold.

As with the constituent sources, the stacked spectra are classified as a detection if SNR $>3$, and the Gaussian fits are used to calculate the flux. Otherwise, a boxcar profile is used to estimate an upper limit for the flux in the same procedure as described in Section \ref{sec:results-gasmass}. The resulting gas properties derived from the stacked spectra are presented in Table \ref{table:stacked}.

\begin{deluxetable*}{chCCCCCCC}[!ht]
  \tabletypesize{\footnotesize}
  \tablecaption{Properties of stacked galaxy images. \label{table:stacked}}
  \tablehead{
    \colhead{Gal} & \nocolhead{Peak Frequency} & \colhead{FWHM} & \colhead{$S_{\mathrm{CO}(3\rightarrow2)}\Delta v$\tablenotemark{a}} & \colhead{RMS} & \colhead{$z_{\mathrm{CO}}$} & \colhead{$L'_{\mathrm{CO}(3\rightarrow2)}$} & \colhead{Delensed $\rm M_{\mathrm{gas}}$} & \colhead{$t_{depl}$}
    \\
    \colhead{ID} & \nocolhead{GHz} & \colhead{${\rm km\,s^{-1}}$} & \colhead{${\rm mJy\,km\,s^{-1}}$} & \colhead{$\times 10^{-4}$ Jy/beam} & \colhead{} & \colhead{$\times 10^{9}~{\rm K\,km\,s^{-1}\,pc^{2}}$} & \colhead{$\times 10^{10}~{\rm M_\odot}$} & \colhead{Gyr}
  }
  \startdata
  \multicolumn{9}{c}{Detections} \\
  \hline
  2 & 88.19\pm0.00 & 279\pm33  & 37.705\pm3.625  & 0.13 & 2.9210\pm0.0004 & 1.58\pm0.15 & 1.36\pm0.13 & 0.14\pm0.01 \\
  5 & 88.19\pm0.02 & 694\pm143 & 83.385\pm12.426 & 0.34 & 2.921\pm0.001 & 3.50\pm0.52 & 3.01\pm0.45 & <0.86 \\
  7 & 88.32\pm0.03 & 379\pm362 & 4.317\pm0.975   & 0.02 & 2.915\pm0.001 & 0.18\pm0.04 & 0.16\pm0.04 & 0.14\pm0.03 \\
  \hline
  \multicolumn{9}{c}{Non-Detections} \\
  \hline
  1 & \nodata & \nodata & <9.429  & 0.08 & \nodata & <0.39 & <0.34 & \nodata \\
  3 & \nodata & \nodata & <50.107 & 0.42 & \nodata & <2.12 & <1.82 & \nodata \\
  4 & \nodata & \nodata & <33.357 & 0.28 & \nodata & <1.41 & <1.21 & \nodata \\
  6 & \nodata & \nodata & <35.120 & 0.29 & \nodata & <1.47 & <1.26 & <0.24 \\
  \enddata
  \tablenotetext{a}{Note that the units are in mJy\,km\,s\inv\, rather than Jy\,km\,s\inv\ as in Table \ref{table:gasprops}.}
  \tablecomments{Properties derived from the \deleted{spectra of the stacked galaxy images}\added{stacked spectra}. Values \deleted{have }are all delensed/intrinsic values as the constituent spectra (eg. Gal 2.a, 2.b, and 2.c) have all been divided by their respective amplifications before the stacking procedure. Stacking was done with a inverse-variance weighted sum. The top section of the table provides values for the galaxies where two or more of the multiple images were detected in \cothreetwo. In these cases, the flux of the stacked spectrum was calculated with a Gaussian fitting procedure as described in Section \ref{sec:results-gasmass}. The rest of the galaxies' fluxes are calculated using a boxcar profile, also described in Section \ref{sec:results-gasmass}. The reported upper limits for these galaxies are $3\sigma$ upper limits.}
\end{deluxetable*}

\Needspace{4\baselineskip}
\section{Discussion}\label{sec:discussion} 

\Needspace{4\baselineskip}
\subsection{Star Formation in High Redshift Galaxies}\label{sec:discussion-starformation}

The molecular gas and the star formation rate are in essence the amount of fuel available to a galaxy and the galaxy's fuel consumption rate. The ratio between the two quantities provides an estimate for the molecular gas depletion time, which is the time that a galaxy will take to consume all of its currently available fuel at its present SFR, assuming there are no other causes of net change in the molecular gas mass.

\begin{figure}[!hb]
  \centering
  \includegraphics[width=1.\linewidth]{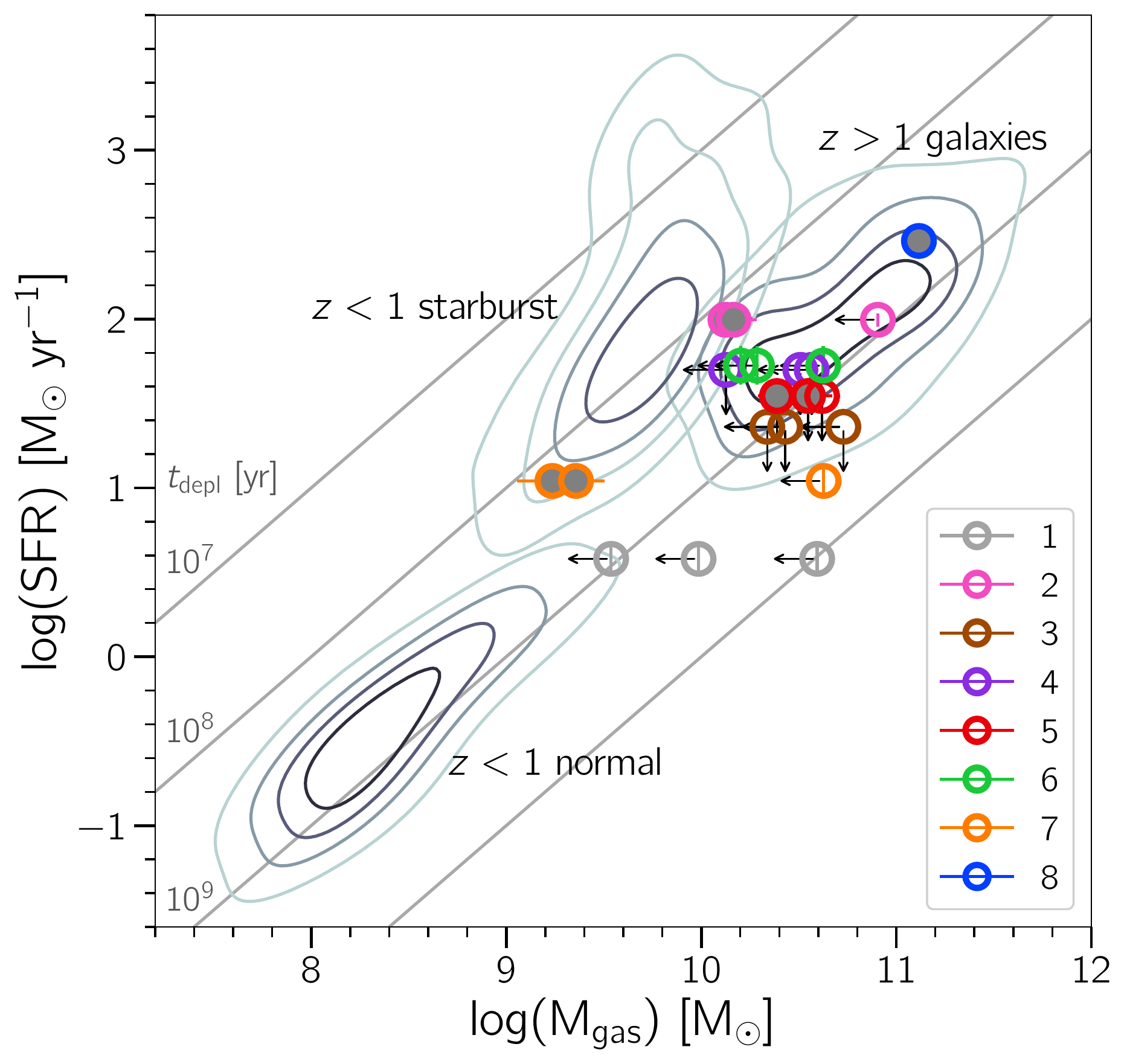}
  \caption{Star formation rate (SFR) as a function of molecular gas mass, plotted in log-log scale\deleted{for the various galaxy populations from Table~\ref{table:compilation}, as annotated in the figure}. All SFRs have been scaled to a \cite{Kroupa_2001} IMF.
    \deleted{The lensed galaxies in this work probe the lower end of the $\mathrm{M_{gas}}-\mathrm{SFR}$ parameter space occupied by existing $z>1$ detections.}
    A compilation of literature measurements listed in Table~\ref{table:compilation} is separated into three galaxy populations as \deleted{labelled}\added{labeled}. These are shown as contours corresponding to levels of constant density (in the probabilistic sense). The contours shown here are at 20\%, 40\%, 60\%, and 80\%, indicating the probability mass below/outside the contour. The darker inner contours indicate the parameter space that a larger number of galaxies occupy.
    Galaxies from this work are plotted separately; detections are plotted with grey face colours, and non-detections are plotted as rings. The ring colour indicates the source galaxy as given by the legend in the lower-right corner.
    For all galaxies except for \added{Gal. }8, there are three points, one for each of the multiple images.
    Diagonal grey lines indicate constant depletion time ($t_{\mathrm{depl}} = \mathrm{M_{gas}} / {\rm SFR}$) as annotated in gray text.
  }
  \label{fig:comparison}
\end{figure}

Figure \ref{fig:comparison} shows the SFR as a function of molecular gas mass for the galaxies from this work, as well as from a compiled sample of over 500 galaxies, the majority of which are CO-detected. This compilation is presented in Table \ref{table:compilation} in Appendix \ref{sec:appendix-compilation}. The compilation includes normal galaxies and starbursts in various environments up to $z=5.3$. The emphasis lies on the high-redshift (using $z>1$ as the cutoff) normal galaxy population where the sources in this work belong. The only included study \deleted{with galaxies that are not CO-detected is \cite{Zavala2019}, which presents dust-detected galaxies}\added{which does not target CO emission is \cite{Zavala2019}, which presents dust continuum observations}. However, this study provides a useful and relatively large sample of high-redshift normal protocluster galaxies as a point of comparison. For starbursts where the infrared (IR) luminosity was given instead of the SFR, Equation 4 from \cite{Kennicutt_1998} was used to calculate the SFR \citep{Ivison2011, Papadopoulos2012, Combes_2013, Magdis_2014, Herrero_Illana_2019}. All SFRs are scaled to a \cite{Kroupa_2001} IMF following the conversion factors provided in \cite{Madau_2014}---to convert from a Salpeter IMF, multiply by $0.67$, and to convert from a Chabrier IMF, multiply by $0.67 / 0.63$.

In Figure \ref{fig:comparison}, we separate the compilation into three populations: $z<1$ normal galaxies, $z<1$ starbursts, and $z>1$ galaxies. We do not distinguish between normal galaxies and starbursts for the $z>1$ population as there is not as clear of a distinction between the two populations.

We find that most of the galaxies from this work lie on the lower end, both in terms of SFR and $\mathrm{M_{gas}}$, of the region of the plot that is most densely populated by other $z>1$ galaxies---clustered around the 1 Gyr line, with $\mathrm{M_{gas}} > 10^{10}~\mathrm{M_\odot}$. However, most of the detections have depletion times closer to $0.1-0.5~{\rm Gyr}$ (see Table \ref{table:gasprops}). The shorter depletion times of these galaxies indicate that they may quench before $z=2.4$, which is roughly 0.5 Gyr after $z=2.9$. By $z=2$, 1 Gyr will have elapsed, and most of the sources will have depleted their molecular gas reservoir.

The outlier of most interest is Gal. 7 because its SFR and gas mass are lower than the bulk of the other ${z>1}$ galaxies by an order of magnitude. \deleted{The fact that it is possible for us to observe this galaxy at all is due to lensing (Gal. 7 is the most strongly lensed source), which allows us to probe intrinsically fainter galaxies.}\added{It is the source at the highest magnification, with the "b" image being magnified by a factor of $45\pm3$, and without lensing, more telescope time than is practical would be required in order to achieve the same SNR as in the presented spectrum.} We also note that all of our limits are upper limits, and the true location of the undetected galaxies in the $\mathrm{M_{gas}}-\mathrm{SFR}$ parameter space is toward the position of Gal. 7, with lower masses and SFRs than other high-redshift detections.

There is a span of two orders of magnitude in the star formation rates of \deleted{our galaxies}\added{the galaxies in this work}, ranging from $3.8_{-0.92}^{+0.98}~{\rm M_\odot\,yr^{-1}}$ for Gal. 1 (Man et al., submitted) to $290\pm90~{\rm M_\odot\,yr^{-1}}$ for Gal. 8 \citep{MacKenzie2014}. This is comparable to what is observed by \cite{Zavala2019}, who find SFRs from as low as $6_{-5}^{+23}~{\rm M_\odot\,yr^{-1}}$ to as high as $612_{-200}^{+280}~{\rm M_\odot\,yr^{-1}}$ in $z\approx2.1$ and $z\approx2.5$ protocluster galaxies. Although \deleted{the number of detections is scarce, and }sensitivity and selection bias play a\deleted{n increasingly} large role \deleted{towards}\added{at} high\deleted{er} redshifts \added{(e.g., selections based on narrow-band filters bias against galaxies with low star-formation activity, and due to the ${\rm M_\ast-SFR}$ relation for typical galaxies, against low mass galaxies \citep{Muldrew_2015})}, \added{the few confirmed} protocluster galaxies at $z>3$ with CO detections form stars much more vigorously than \deleted{our sample, with SFRs exceeding $6000~{\rm M_\odot\,yr^{-1}}$ for a single protocluster \citep{Miller2018, Oteo2018}.}\added{the galaxies in this work. For example, the protoclusters from \cite{Miller2018} and \cite{Oteo2018} have total SFRs exceeding $6000~{\rm M_\odot\,yr^{-1}}$.}

The large range of SFRs at $z=1$ to $3$---and possibly at even higher redshifts, but there are too few detections to say with certainty---suggest that protocluster galaxies are quenched during the $z=1$ to $3$ period. During this period, we concurrently see starburst galaxies with high SFRs and post-starburst galaxies with nearly no star formation. It is possible that the quenching of protocluster galaxies is aided by the dense environments---promoting more interactions which could lead to quenching \citep{Lotz2013}---in which these galaxies reside. If indeed galaxies in dense environments are more rapidly quenched than field galaxies, a high fraction of quiescent galaxies would be observed in dense environments at $z<1$. Existing studies on the impact of the environment on quenching support this \citep{Cooper_2007, Wang2018, Strazzullo_2019, Castignani2020}. Preferential quenching in dense environments would also mean that the more strongly star forming galaxies would be in the field at later cosmic epochs. This is supported by the inversion in the density-SFR relation at $z\approx1$, showing that locally, galaxies with high SFRs are preferentially found in isolated environments \citep{Dressler_1980, Gomez_2003, Cooper_2007, Elbaz_2007, Tran_2010, Popesso_2015a, Popesso_2015b}.

\Needspace{4\baselineskip}
\subsection{Dependence of Depletion Time on Redshift}\label{sec:discussion-redshift}

\begin{figure}[!ht]
  \centering
  \includegraphics[width=1\linewidth]{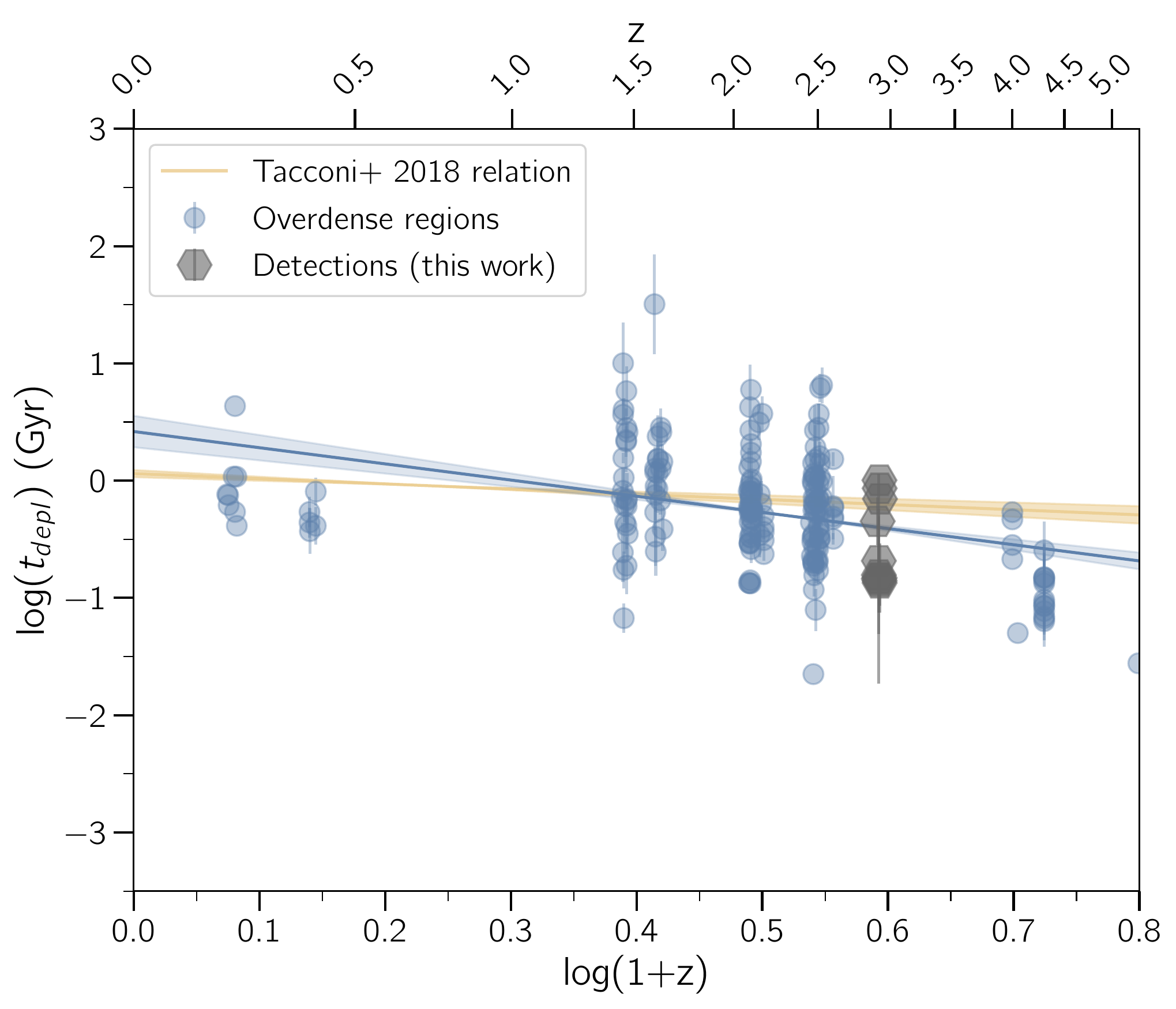}
  \caption{Depletion time $t_{\mathrm{depl}} = \mathrm{M_{gas}} / \text{SFR}$ in Gyr as a function of redshift\deleted{ for galaxies from Table \ref{table:compilation}}, plotted in log-log scale. \deleted{Galaxies in dense environments appear to have longer depletion times that evolve more strongly with redshift than their counterparts in the field. However, more observations at $z > 3$ are required to confirm these findings. We use a different colour for the galaxies from this work to make them easier to identify. }The \deleted{teal}\added{blue} line and shading indicate the line of best fit and the corresponding $1\,\sigma$ uncertainty from 10,000 bootstrapped samples of galaxies in overdense environments (see Table \ref{table:compilation}). \deleted{The purple points are field galaxies. }The yellow line is the field scaling relation from \cite{Tacconi_2018}, where only the impact of redshift is considered. \added{The depletion time here is given by $\log(t_{\rm depl}) = (0.06\pm0.03) - (0.44\pm0.13) \log(1+z) - (0.43\pm0.03) \log(\delta{\rm MS}) + (0.17\pm0.04) (\log{\rm M_\ast} - 10.7)$, and we assume normalization factors of $\delta \text{MS}=1$ and $\log \text{M}_\ast = 10.7$ so that redshift is the only factor influencing the depletion time}.}
  \label{fig:tdepl}
\end{figure}

Figure \ref{fig:tdepl} shows the redshift-evolution of the depletion time, distinguished by the environment. \added{The data are from compiled galaxies in Table \ref{table:compilation}, where studies are marked as "overdense" if they provide measurements of (proto-)cluster galaxies.} The blue line is the fit to galaxies in overdense environments, and the blue shading is the $1\,\sigma$ confidence interval on that fit. The fit and uncertainty are calculated from the sample means and the standard errors of the line parameters as determined by the Levenberg-Marquardt algorithm for 10,000 bootstrapped samples \citep{Levenberg1944, Efron_1979}. \deleted{The purple points are field galaxies from the compilation. }Also plotted in yellow for comparison is the field scaling relation for CO measurements from \cite{Tacconi_2018}.\added{ The depletion time here is given by $\log(t_{\rm depl}) = (0.06\pm0.03) - (0.44\pm0.13) \log(1+z) - (0.43\pm0.03) \log(\delta{\rm MS}) + (0.17\pm0.04) (\log{\rm M_\ast} - 10.7)$. This scaling relation is for CO measurements, and is informed by 667 galaxies. }The normalization factors \deleted{assumed}\added{we assume} for this relation are $\delta \text{MS}=1$ and $\log \text{M}_\ast = 10.7$, where $\delta \text{MS} = \text{sSFR} / \text{sSFR}(\text{MS}, z, \rm{M}_\ast)$ is the offset from the star formation main sequence (MS). \added{These are chosen so that the relation is only dependent on redshift. Although the depletion time depends on main-sequence offset and stellar mass (the $C_t$ and $D_t$ coefficients in \citealt{Tacconi_2018}), we do not consider these factors due to a lack of adequate data for accurate measurements.}\deleted{this means that the relation is only dependent on redshift.}

Looking at the evolution of the depletion time of galaxies\deleted{ in overdense environments with redshift}, the \deleted{data suggest}\added{fit suggests} that \deleted{the $z>2.5$ protocluster }\added{ high-redshift }galaxies \deleted{are at}\added{have} shorter depletion times \deleted{compared to}\added{than} lower redshift \deleted{cluster }galaxies.
\deleted{We find that the depletion time for galaxies from the compilation scales with redshift as $t_{\mathrm{depl}} \propto (1+z)^{-1.37\pm0.26}$.}\added{For galaxies in overdense environments, the slope of the best-fit line is $t_{\mathrm{depl}} \propto (1+z)^{-1.37\pm0.26}$.}
\added{We note here that there may be biases against long depletion times observationally and/or in the sample we have compiled, which cannot be accounted for in the line fitting.}
We compare this to the \cite{Tacconi_2018} field relation of $t_{\mathrm{depl}} \propto (1+z)^{-0.44\pm0.13}$\deleted{ for CO measurements}\added{, which is less steep}, \deleted{showing}\added{suggesting} that \deleted{depletion times for galaxies in overdense environments evolve more strongly with redshift than depletion times for field galaxies.}\added{while depletion time decreases with increasing redshift for all galaxies, the evolution is stronger for galaxies in overdense environments.}
\cite{Darvish_2018} also investigated the redshift evolution of depletion time for galaxies in various environmental densities up to $z\approx3.5$ and found that the depletion time decreases with increasing redshift. However, they did not find evidence for a dependence of the \added{redshift-depletion time} relation on environmental density.

\Needspace{4\baselineskip}
\subsection{Dependence of Depletion Time on Environmental Density}\label{sec:discussion-environment}

\deleted{Up to redshifts of $z=2.5$, \cite{Rudnick_2017}, \cite{Hayashi2018}, and \cite{Tadaki2019} find that cluster galaxies (in dense environments) have longer }\added{In line with other studies, we find that galaxies in overdense environments at ${z<2}$ have depletion times that are roughly consistent with \citep{Noble2017, Rudnick_2017, Darvish_2018} the depletion times of }coeval field galaxies\added{, as determined by field scaling relations (from \citealt{Tacconi2013, Tacconi_2018} or \citealt{Genzel2015})}. This is in spite of the fact that high-redshift cluster galaxies have similar or larger gas fractions compared to coeval field galaxies \citep{Noble2017, Hayashi2018}, suggesting that galaxies in overdense environments are less efficient at converting gas into stars. \deleted{A possible explanation for this is that processes such as ram-pressure stripping and strangulation, which occur preferentially in overdense environments, are preventing the cooling of gas, and thus is preventing star formation \citep{Mo_2010}. }\added{We also note that the stellar mass is likely an important factor in the efficacy of these environmental processes.}

\added{From the fit shown in Figure \ref{fig:tdepl} (which includes many starburst galaxies), galaxies in overdense environments seem to have shorter depletion times than field galaxies at $z>2.5$. However, this fit is informed by mostly $z<2.5$ galaxies, many of which are starbursts. If we analyze the small subset of CO-detected galaxies at $z>2.5$ with normal SFRs, this becomes unclear. In the \cite{Tacconi_2018} sample, only four detections belong to this subset: one non-lensed Lyman break galaxy from \cite{Magdis_2012} (M23), and three lensed galaxies from \cite{Saintonge_2013} (cB58, 8:00arc, and Eye). The mean depletion time for these four galaxies is $0.21\pm0.04~{\rm Gyr}$. From the compilation, there are 23 galaxies \added{with normal SFRs }in overdense environments \deleted{in this subset }with known depletion times---19 from \cite{Tadaki2019} and \cite{Wang2018}, and 5 from this work (Gals. 2.b, 2.c, 7.a, 7.b, and 8). The mean depletion time for these 23 galaxies is longer, at $1.14\pm0.14~{\rm Gyr}$. Based on this very small sample, it would seem then, that even at high redshifts, the depletion times of galaxies in overdense environments are longer than those of galaxies in field environments.}

The primary problem preventing consensus in this discussion is the scarcity of CO-detected \deleted{protocluster }galaxies at $z>2.5$ with normal star formation rates\added{, both in the field and in overdense environments}. Current studies of \added{molecular gas, particularly those with CO, at }high-redshift \deleted{protocluster galaxies }are primarily composed of massive starbursts \citep[e.g.,][]{Riechers_2010a, Bothwell_2013, Oteo2018, Miller2018}. More observations of high-redshift normal galaxies are needed in order to better constrain the relationship between redshift and depletion time, and to identify the impact of the environment on that relationship.

\Needspace{4\baselineskip}
\subsection{Star-Gas Offset}\label{sec:discussion-offset}

\added{We find \hst-ALMA spatial offsets of $0.14-0.33$\arcsec\ on the source plane, corresponding to $1.1-2.6~{\rm kpc}$ at ${z=2.9}$. Such offsets are}\deleted{The \hst-ALMA spatial offset observed in our sources is} not particularly surprising given the conditions of our sources, and \deleted{has}\added{have} also been observed in other high-redshift galaxies \citep{Carilli_2010, Daddi_2010, Hodge_2013}. Galaxies at high redshift are dustier than their local counterparts \citep{Hodge2020}. It is conceivable that in our case the dust is co-spatial with the CO emissions, and obscures a large portion of UV emission from the galaxies. The spatial offset would only be exacerbated by interactions between galaxies---which are more likely to happen in overdense environments---as they could cause gas and dust to be flung out in a way that it becomes extended and/or offset relative to the stars in \deleted{the }galaxies \citep{Zwicky1956, Barnes1992}.

The multiply-imaged nature of the lensed sources lends further evidence that the observed offset is intrinsic rather than due to the astrometry. Consider Gals. 2.b and 2.c, two of the most strongly detected sources with SNRs of 6.4 and 5.8 respectively. Figure \ref{fig:hst} in this work and Figure 1 from \cite{MacKenzie2014} show that the offset is mirrored across the critical line. \replaced{This is further corroborated by the CO rotation, seen in the moment one maps, also being}{The velocity gradients in the moment one maps, which are calculated with the equation
  \begin{align}
    M_1 = \frac{\int I v dv}{\int I dv},
  \end{align}
where $I$ is the intensity of the pixel value in Jy and $v$ is the velocity of the channel in $\rm{km\,s^{-1}}$, and the integration goes across the $-200~{\rm km\,s^{-1}}$ to $200~{\rm km\,s^{-1}}$ velocity channels for each pixel in the optimal extraction region (see Section \ref{sec:results-specextraction}), }are also mirrored across the critical line (see Figure \ref{fig:2bc_moment1}).\added{ Moment one maps of all detected galaxies are given in Appendix \ref{sec:appendix-m1}. }A closer look at the \hst\ images also reveals \deleted{a mirrored color gradient:}\added{that} the CO emission is co-spatial with \deleted{red, diffuse}\added{diffuse, red} blobs of dust near each source in all the images. If in fact the offset was due to astrometry, we would not expect to see any of these things; rather, the offset would be of a similar magnitude and direction in all the sources.

\begin{figure}[!ht]
  \centering
  \includegraphics[width=1\linewidth]{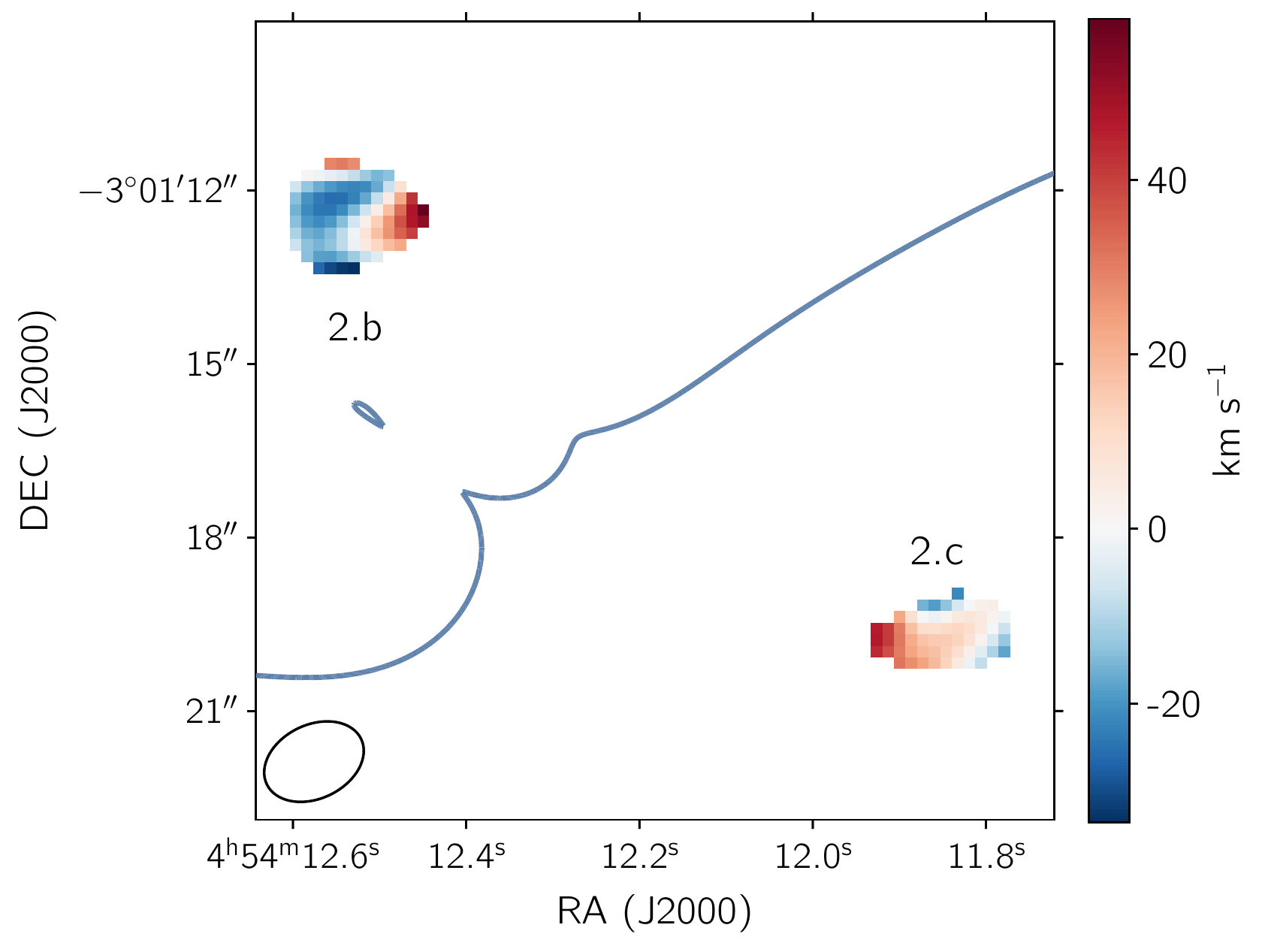}
  \caption{Moment one map of Gal. 2.b (left) and 2.c (right), as created from data cubes masked to only retain the optimal extraction region (see Section \ref{sec:results-specextraction} for details)\deleted{. The sources are not well resolved, and thus it is difficult to make conclusive statements about the gas dynamics.}\added{ of each source, plotted together with the critical line of the lens model.} A velocity gradient of approximately $80~{\rm km\,s^{-1}}$ is present in both sources, and the directional alignment of the gradient in Gal. 2.b appears to be mirrored across the critical line relative to that \deleted{in}\added{of} Gal. 2.c. The velocities are given relative to the systematic velocity of the respective sources. The black elliptical ring in the bottom left corner indicate the ALMA beam.}
  \label{fig:2bc_moment1}
\end{figure}

\Needspace{4\baselineskip}
\subsection{Systematic Uncertainties in the Gas Mass}\label{sec:discussion-uncertainties}

The measurement of the molecular gas mass relies on two factors, both of which carry significant systematic uncertainties. The first is the ratio between the strengths of the different CO rotational transitions, or the CO SLED. The \cothreetwo\ transition requires more energy to be excited than the ground-state transition, and thus it traces denser clumps of gas with higher temperatures \citep{Riechers_2007}. A conversion factor $r_{3\rightarrow2 / 1\rightarrow0} = L^{\prime}_{\mathrm{CO(3\rightarrow2)}} / L^\prime_{\mathrm{CO(1\rightarrow0)}}$ needs to be used in order to infer the \coonezero\ luminosity. However, a wide range of values have been adopted in literature, ranging from $0.4-1.15$ in $z\approx2.5$ galaxies \citep{Ivison2011, Gomez-Guijarro_2019}. This conversion factor is also increasingly impacted by the cosmic microwave background (CMB) at high redshifts \citep{da_Cunha_2013}; the elevated CMB in the early Universe reduces the brightness contrast of CO line emissions against background emissions \citep{Zhang_2016}. Across all redshifts, the value most commonly adopted or calculated by the studies in our compilation targeting that transition is 0.5. This is the value we use for our calculations.

The second major uncertainty is the CO-to-H$_{2}$ conversion factor, $\alpha_{\mathrm{CO}}$. This conversion factor is dependent on many physical properties like the metallicity and the pressure of the interstellar medium (ISM) \citep{Genzel_2012, Carilli2013, Dessauges_Zavadsky_2017}. This is particularly problematic at high redshifts and in low-mass galaxies where there is a rapid metallicity evolution \citep{Tan_2013}. Even within the Milky Way and nearby spirals where resolved observations are easier to obtain and ISM properties are better understood, $\alpha_{\mathrm{CO}}$ is not certain, with \cite{Bolatto_2013} finding a factor of two scatter. For extragalactic sources, it is much more difficult to \deleted{apply these methods} to determine $\alpha_{\mathrm{CO}}$. For Milky Way-like galaxies, a value of $\alpha_{\mathrm{CO}}=4.36~{\rm M_\odot\,(K\,km\,s^{-1}\,pc^{-2})}$ is commonly used, and for (U)LIRGs and SMGs, $\alpha_{\mathrm{CO}}=0.8$ is commonly used \citep{Downes_1998, Dessauges_Zavadsky_2015, Noble2019}. If metallicity measurements are available, it may be beneficial to adopt a CO line intensity and metallicity dependent $\alpha_{\mathrm{CO}}$ factor from the best-fitting function in \cite{Narayanan_2012} rather than the conventional bimodal $\alpha_{\mathrm{CO}}$ values, as this has been shown to significantly reduce scatter in star formation relations.

Together, the CO SLED and $\alpha_{\mathrm{CO}}$ factors introduce close to an order of magnitude in systematic uncertainty into \deleted{our results}{the ${\rm M_{\rm gas}}$ and $t_{\rm depl}$ derived in this work}.

\Needspace{4\baselineskip}
\subsection{Future Prospects}\label{sec:discussion-future}

Future observations of these sources and other lensed high-redshift galaxies could take many avenues. \deleted{Observations with the ground-state CO transition}\added{Future instruments like the Next Generation Very Large Array (ngVLA) and ALMA Band 1 will enable large, multi-transition CO surveys of more representative galaxies at $z>2$, which} would open up the possibility to model the CO SLED (e.g., \cite{Narayanan2014}). Higher spatial resolution observations would allow for virial mass estimates that could be used to independently constrain $\alpha_{\mathrm{CO}}$,
and to reconstruct the gas distribution on the source plane (e.g., \citealp{Brewer_2006, Riechers_2007, Vegetti_2009, Spingola_2020}).

It is conceivable that the overdense environment is more conducive to mergers and active galactic nuclei (AGN) activity. Radio counterparts to some of the sources in this work have been identified by \cite{Berciano_Alba_2010} (e.g., "RJ Bright" near Gal 1.b and the three multiple images of the "E" system near the three multiple images of Gals. 5 and 6).
These could be indications of the existence of potential mergers and/or AGN in the galaxies. We examined the spectra for broader CO components, but the ALMA data at hand are not sufficiently deep to identify signatures of AGN-driven molecular gas outflows. Future investigations into the presence of AGN could shed light on its impact on star formation in protocluster galaxies.

Follow-up \added{rest-frame optical} imaging could also aim to \deleted{obtain}\added{provide} information about the stellar mass of the \deleted{sources in this work}\added{galaxies of this ${z=2.9}$ group}. That would allow for the calculation of gas fractions in these galaxies, which could lead to further understanding of the impact of the environment on molecular gas. Stellar masses could also be used to compare the sources to the star formation main sequence, allowing for a more comprehensive understanding of star formation in these galaxies.

\Needspace{4\baselineskip}
\section{Conclusions}\label{sec:conclusion} 

In this paper, we have presented a study of a gravitationally lensed group of galaxies at $z=2.9$. With a compact ALMA configuration, we target the \cothreetwo\ emission line and spectroscopically confirm the redshifts of seven images of four members. These molecular gas mass measurements represent the highest redshift CO detections in dense environments for galaxies that are not undergoing extreme starburst. We compile a diverse sample of over 500 galaxies up to $z=5.3$ in order to place our results in context. We find that the galaxies in this work are consistent in molecular gas mass and SFR with the high-redshift ($z>1$) normal population from our compilation. The inferred molecular gas masses of our detections span a range of $(0.2-13.1)\times 10^{10}~{\rm M_{\odot}}$.

We also investigated the impact of the environment on the depletion time of galaxies. The sources in this work are among the highest redshift CO-detected galaxies with normal star formation rates in an \deleted{dense}\added{overdense} environment, and extend the discussion about the impact of the environment beyond $z=2.5$. At the same time, our results highlight the challenge of constraining molecular gas relations for normal galaxies at high-redshift due to the scarcity of observations. \deleted{We tentatively find that galaxies in overdense environments have a longer depletion time than field galaxies, as well as a steeper evolution with redshift.}\added{We find that at high redshifts, galaxies tend to have shorter depletion times than those at low redshifts, and that this depletion time evolution is steeper for galaxies in overdense environments than for field galaxies. At $z>2.5$, we find tentative evidence that those in overdense environments have shorter depletion times than coeval field galaxies, but this does not seem to hold when considering only CO-detected galaxies with normal star formation rates.} \deleted{However, the large scatter in the depletion times, as well as }The lack of CO detections of normal \deleted{protocluster galaxies past $z>3$}\added{galaxies at $z>2.5$}\deleted{,} makes it difficult to conclusively determine the impact of the environment on the depletion time. Further investigation into cluster evolution will be needed in order to fully understand how early galaxies in dense environments evolve.

\Needspace{4\baselineskip}
\section*{Acknowledgements}\label{sec:acknowledgements}
\added{We thank the anonymous referee for a careful reading of this manuscript and for providing helpful suggestions. }The Dunlap Institute is funded through an endowment established by the David Dunlap family and the University of Toronto. The University of Toronto operates on the traditional land of the Huron-Wendat, the Seneca, and most recently, the Mississaugas of the Credit River; J.S. and A.M. are grateful to have the opportunity to work on this land.
This paper makes use of data from ALMA program ADS/JAO.ALMA\#2017.1.00616.S. ALMA is a partnership of ESO (representing its member states), NSF (USA) and NINS (Japan), together with NRC (Canada) and NSC and ASIAA (Taiwan) and KASI (Republic of Korea), in cooperation with the Republic of Chile. The Joint ALMA Observatory is operated by ESO, AUI/NRAO and NAOJ.
J.Z. acknowledges support from ANR grant ANR-17-CE31-0017 (3DGasFlows).
We acknowledge Sune Toft, Harald Ebeling, Desika Narayanan, and Justin Spilker for their contributions to earlier stages of this work.

\software{NumPy \citep{van_der_Walt_2011}, spectral-cube \citep{spectral-cube}, Matplotlib \citep{Hunter_2007}, Astropy \citep{Astropy_2013, Astropy_2018}, seaborn \citep{waskom2020seaborn}, SciPy \citep{SciPy_2020}, CASA \citep{McMullin2007}, grizli \citep{Brammer2019}}

\section*{Author Contributions}\label{sec:contributions}

J.S. performed the data analysis, created the figures, wrote most of the paper, and made revisions under the supervision of A.M. A.M. designed the study, led the writing of the ALMA proposal, configured and processed the ALMA observations, and wrote Section \ref{sec:observations}. J.Z. contributed to the design of the study. Z.Z. contributed to the ALMA imaging. J.Z., Z.Z., and M.S. provided feedback on the ALMA proposal. G.B. processed the \hst\ images. \added{J.R. performed the lens model inversion.} All authors participated in scientific discussions and provided comments which led to revisions of the paper \added{both} prior to \added{and following} submission.

\bibliography{sources}

\begin{thebibliography}{}
\expandafter\ifx\csname natexlab\endcsname\relax\def\natexlab#1{#1}\fi
\providecommand{\url}[1]{\href{#1}{#1}}
\providecommand{\dodoi}[1]{doi:~\href{http://doi.org/#1}{\nolinkurl{#1}}}
\providecommand{\doeprint}[1]{\href{http://ascl.net/#1}{\nolinkurl{http://ascl.net/#1}}}
\providecommand{\doarXiv}[1]{\href{https://arxiv.org/abs/#1}{\nolinkurl{https://arxiv.org/abs/#1}}}

\bibitem[{{Barnes} \& {Hernquist}(1992)}]{Barnes1992}
{Barnes}, J.~E., \& {Hernquist}, L. 1992, \araa, 30, 705,
  \dodoi{10.1146/annurev.aa.30.090192.003421}

\bibitem[{{Berciano Alba} {et~al.}(2007){Berciano Alba}, {Garrett}, {Koopmans},
  \& {Wucknitz}}]{Berciano_Alba_2007}
{Berciano Alba}, A., {Garrett}, M.~A., {Koopmans}, L.~V.~E., \& {Wucknitz}, O.
  2007, \aap, 462, 903, \dodoi{10.1051/0004-6361:20065223}

\bibitem[{Berciano~Alba {et~al.}(2010)Berciano~Alba, Koopmans, Garrett,
  Wucknitz, \& Limousin}]{Berciano_Alba_2010}
Berciano~Alba, A., Koopmans, L. V.~E., Garrett, M.~A., Wucknitz, O., \&
  Limousin, M. 2010, Astronomy and Astrophysics, 509, A54,
  \dodoi{10.1051/0004-6361/200912903}

\bibitem[{Blanton \& Moustakas(2009)}]{Blanton_2009}
Blanton, M.~R., \& Moustakas, J. 2009, Annual Review of Astronomy and
  Astrophysics, 47, 159–210, \dodoi{10.1146/annurev-astro-082708-101734}

\bibitem[{Bolatto {et~al.}(2013)Bolatto, Wolfire, \& Leroy}]{Bolatto_2013}
Bolatto, A.~D., Wolfire, M., \& Leroy, A.~K. 2013, Annual Review of Astronomy
  and Astrophysics, 51, 207–268, \dodoi{10.1146/annurev-astro-082812-140944}

\bibitem[{Borys {et~al.}(2004)Borys, Chapman, Donahue, Fahlman, Halpern, Kneib,
  Newbury, Scott, \& Smith}]{Borys_2004}
Borys, C., Chapman, S., Donahue, M., {et~al.} 2004, Monthly Notices of the
  Royal Astronomical Society, 352, 759–767,
  \dodoi{10.1111/j.1365-2966.2004.07982.x}

\bibitem[{{Bothwell} {et~al.}(2013){Bothwell}, {Smail}, {Chapman}, {Genzel},
  {Ivison}, {Tacconi}, {Alaghband-Zadeh}, {Bertoldi}, {Blain}, {Casey}, {Cox},
  {Greve}, {Lutz}, {Neri}, {Omont}, \& {Swinbank}}]{Bothwell_2013}
{Bothwell}, M.~S., {Smail}, I., {Chapman}, S.~C., {et~al.} 2013, \mnras, 429,
  3047, \dodoi{10.1093/mnras/sts562}

\bibitem[{{Brammer}(2019)}]{Brammer2019}
{Brammer}, G. 2019, {Grizli: Grism redshift and line analysis software}.
\newblock \doeprint{1905.001}

\bibitem[{Brewer \& Lewis(2006)}]{Brewer_2006}
Brewer, B.~J., \& Lewis, G.~F. 2006, The Astrophysical Journal, 637, 608–619,
  \dodoi{10.1086/498409}

\bibitem[{Carilli \& Walter(2013)}]{Carilli2013}
Carilli, C., \& Walter, F. 2013, Annual Review of Astronomy and Astrophysics,
  51, 105, \dodoi{10.1146/annurev-astro-082812-140953}

\bibitem[{Carilli {et~al.}(2010)Carilli, Daddi, Riechers, Walter, Weiss,
  Dannerbauer, Morrison, Wagg, Davé, Elbaz, \& et~al.}]{Carilli_2010}
Carilli, C.~L., Daddi, E., Riechers, D., {et~al.} 2010, The Astrophysical
  Journal, 714, 1407–1417, \dodoi{10.1088/0004-637x/714/2/1407}

\bibitem[{Castignani {et~al.}(2020)Castignani, Combes, Salom{\'{e}}, \&
  Freundlich}]{Castignani2020}
Castignani, G., Combes, F., Salom{\'{e}}, P., \& Freundlich, J. 2020, Astronomy
  {\&} Astrophysics, 635, A32, \dodoi{10.1051/0004-6361/201936148}

\bibitem[{Chambers {et~al.}(2019)Chambers, Magnier, Metcalfe, Flewelling,
  Huber, Waters, Denneau, Draper, Farrow, Finkbeiner, Holmberg, Koppenhoefer,
  Price, Rest, Saglia, Schlafly, Smartt, Sweeney, Wainscoat, Burgett, Chastel,
  Grav, Heasley, Hodapp, Jedicke, Kaiser, Kudritzki, Luppino, Lupton, Monet,
  Morgan, Onaka, Shiao, Stubbs, Tonry, White, Bañados, Bell, Bender, Bernard,
  Boegner, Boffi, Botticella, Calamida, Casertano, Chen, Chen, Cole, Deacon,
  Frenk, Fitzsimmons, Gezari, Gibbs, Goessl, Goggia, Gourgue, Goldman, Grant,
  Grebel, Hambly, Hasinger, Heavens, Heckman, Henderson, Henning, Holman, Hopp,
  Ip, Isani, Jackson, Keyes, Koekemoer, Kotak, Le, Liska, Long, Lucey, Liu,
  Martin, Masci, McLean, Mindel, Misra, Morganson, Murphy, Obaika, Narayan,
  Nieto-Santisteban, Norberg, Peacock, Pier, Postman, Primak, Rae, Rai, Riess,
  Riffeser, Rix, Röser, Russel, Rutz, Schilbach, Schultz, Scolnic, Strolger,
  Szalay, Seitz, Small, Smith, Soderblom, Taylor, Thomson, Taylor, Thakar,
  Thiel, Thilker, Unger, Urata, Valenti, Wagner, Walder, Walter, Watters,
  Werner, Wood-Vasey, \& Wyse}]{Chambers2019}
Chambers, K.~C., Magnier, E.~A., Metcalfe, N., {et~al.} 2019, The Pan-STARRS1
  Surveys.
\newblock \doarXiv{1612.05560}

\bibitem[{Cicone {et~al.}(2017)Cicone, Bothwell, Wagg, Møller, De~Breuck,
  Zhang, Martín, Maiolino, Severgnini, Aravena, \& et~al.}]{Cicone_2017}
Cicone, C., Bothwell, M., Wagg, J., {et~al.} 2017, Astronomy {\&} Astrophysics,
  604, A53, \dodoi{10.1051/0004-6361/201730605}

\bibitem[{Ciesla {et~al.}(2020)Ciesla, Béthermin, Daddi, Richard, Diaz-Santos,
  Sargent, Elbaz, Boquien, Wang, Schreiber, \& et~al.}]{Ciesla_2020}
Ciesla, L., Béthermin, M., Daddi, E., {et~al.} 2020, Astronomy {\&}
  Astrophysics, 635, A27, \dodoi{10.1051/0004-6361/201936727}

\bibitem[{Combes {et~al.}(2013)Combes, García-Burillo, Braine, Schinnerer,
  Walter, \& Colina}]{Combes_2013}
Combes, F., García-Burillo, S., Braine, J., {et~al.} 2013, Astronomy {\&}
  Astrophysics, 550, A41, \dodoi{10.1051/0004-6361/201220392}

\bibitem[{Combes {et~al.}(2012)Combes, Rex, Rawle, Egami, Boone, Smail,
  Richard, Ivison, Gurwell, Casey, \& et~al.}]{Combes_2012}
Combes, F., Rex, M., Rawle, T.~D., {et~al.} 2012, Astronomy {\&} Astrophysics,
  538, L4, \dodoi{10.1051/0004-6361/201118750}

\bibitem[{Cooper {et~al.}(2007)Cooper, Newman, Weiner, Yan, Willmer, Bundy,
  Coil, Conselice, Davis, Faber, \& et~al.}]{Cooper_2007}
Cooper, M.~C., Newman, J.~A., Weiner, B.~J., {et~al.} 2007, Monthly Notices of
  the Royal Astronomical Society, 383, 1058–1078,
  \dodoi{10.1111/j.1365-2966.2007.12613.x}

\bibitem[{Coppin {et~al.}(2007)Coppin, Swinbank, Neri, Cox, Smail, Ellis,
  Geach, Siana, Teplitz, Dye, \& et~al.}]{Coppin_2007}
Coppin, K. E.~K., Swinbank, A.~M., Neri, R., {et~al.} 2007, The Astrophysical
  Journal, 665, 936–943, \dodoi{10.1086/519789}

\bibitem[{Cybulski {et~al.}(2016)Cybulski, Yun, Erickson, De~la Luz, Narayanan,
  Montaña, Sánchez, Zavala, Zeballos, Chung, \& et~al.}]{Cybulski_2016}
Cybulski, R., Yun, M.~S., Erickson, N., {et~al.} 2016, Monthly Notices of the
  Royal Astronomical Society, 459, 3287–3306, \dodoi{10.1093/mnras/stw798}

\bibitem[{da~Cunha {et~al.}(2013)da~Cunha, Groves, Walter, Decarli, Weiss,
  Bertoldi, Carilli, Daddi, Elbaz, Ivison, \& et~al.}]{da_Cunha_2013}
da~Cunha, E., Groves, B., Walter, F., {et~al.} 2013, The Astrophysical Journal,
  766, 13, \dodoi{10.1088/0004-637x/766/1/13}

\bibitem[{Daddi {et~al.}(2010)Daddi, Bournaud, Walter, Dannerbauer, Carilli,
  Dickinson, Elbaz, Morrison, Riechers, Onodera, \& et~al.}]{Daddi_2010}
Daddi, E., Bournaud, F., Walter, F., {et~al.} 2010, The Astrophysical Journal,
  713, 686–707, \dodoi{10.1088/0004-637x/713/1/686}

\bibitem[{Daddi {et~al.}(2015)Daddi, Dannerbauer, Liu, Aravena, Bournaud,
  Walter, Riechers, Magdis, Sargent, Béthermin, \& et~al.}]{Daddi_2015}
Daddi, E., Dannerbauer, H., Liu, D., {et~al.} 2015, Astronomy {\&}
  Astrophysics, 577, A46, \dodoi{10.1051/0004-6361/201425043}

\bibitem[{Danielson {et~al.}(2010)Danielson, Swinbank, Smail, Cox, Edge, Weiss,
  Harris, Baker, De~Breuck, Geach, \& et~al.}]{Danielson_2010}
Danielson, A. L.~R., Swinbank, A.~M., Smail, I., {et~al.} 2010, Monthly Notices
  of the Royal Astronomical Society, no–no,
  \dodoi{10.1111/j.1365-2966.2010.17549.x}

\bibitem[{Dannerbauer {et~al.}(2017)Dannerbauer, Lehnert, Emonts, Ziegler,
  Altieri, De~Breuck, Hatch, Kodama, Koyama, Kurk, \&
  et~al.}]{Dannerbauer_2017}
Dannerbauer, H., Lehnert, M.~D., Emonts, B., {et~al.} 2017, Astronomy {\&}
  Astrophysics, 608, A48, \dodoi{10.1051/0004-6361/201730449}

\bibitem[{Darvish {et~al.}(2018)Darvish, Scoville, Martin, Mobasher,
  Diaz-Santos, \& Shen}]{Darvish_2018}
Darvish, B., Scoville, N.~Z., Martin, C., {et~al.} 2018, The Astrophysical
  Journal, 860, 111, \dodoi{10.3847/1538-4357/aac836}

\bibitem[{Dessauges-Zavadsky {et~al.}(2015)Dessauges-Zavadsky, Zamojski,
  Schaerer, Combes, Egami, Swinbank, Richard, Sklias, Rawle, Rex, \&
  et~al.}]{Dessauges_Zavadsky_2015}
Dessauges-Zavadsky, M., Zamojski, M., Schaerer, D., {et~al.} 2015, Astronomy
  {\&} Astrophysics, 577, A50, \dodoi{10.1051/0004-6361/201424661}

\bibitem[{Dessauges-Zavadsky {et~al.}(2017)Dessauges-Zavadsky, Zamojski,
  Rujopakarn, Richard, Sklias, Schaerer, Combes, Ebeling, Rawle, Egami, \&
  et~al.}]{Dessauges_Zavadsky_2017}
Dessauges-Zavadsky, M., Zamojski, M., Rujopakarn, W., {et~al.} 2017, Astronomy
  {\&} Astrophysics, 605, A81, \dodoi{10.1051/0004-6361/201628513}

\bibitem[{Downes \& Solomon(1998)}]{Downes_1998}
Downes, D., \& Solomon, P.~M. 1998, The Astrophysical Journal, 507, 615–654,
  \dodoi{10.1086/306339}

\bibitem[{{Dressler}(1980)}]{Dressler_1980}
{Dressler}, A. 1980, \apj, 236, 351, \dodoi{10.1086/157753}

\bibitem[{Efron(1979)}]{Efron_1979}
Efron, B. 1979, The Annals of Statistics, 7, 1.
\newblock \url{http://www.jstor.org/stable/2958830}

\bibitem[{{Egami} {et~al.}(2010){Egami}, {Rex}, {Rawle},
  {P{\'e}rez-Gonz{\'a}lez}, {Richard}, {Kneib}, {Schaerer}, {Altieri},
  {Valtchanov}, {Blain}, {Fadda}, {Zemcov}, {Bock}, {Boone}, {Bridge},
  {Clement}, {Combes}, {Dessauges-Zavadsky}, {Dowell}, {Ilbert}, {Ivison},
  {Jauzac}, {Lutz}, {Metcalfe}, {Omont}, {Pell{\'o}}, {Pereira}, {Rieke},
  {Rodighiero}, {Smail}, {Smith}, {Tramoy}, {Walth}, {van der Werf}, \&
  {Werner}}]{Egami2010}
{Egami}, E., {Rex}, M., {Rawle}, T.~D., {et~al.} 2010, \aap, 518, L12,
  \dodoi{10.1051/0004-6361/201014696}

\bibitem[{Elbaz {et~al.}(2007)Elbaz, Daddi, Le~Borgne, Dickinson, Alexander,
  Chary, Starck, Brandt, Kitzbichler, MacDonald, \& et~al.}]{Elbaz_2007}
Elbaz, D., Daddi, E., Le~Borgne, D., {et~al.} 2007, Astronomy {\&}
  Astrophysics, 468, 33–48, \dodoi{10.1051/0004-6361:20077525}

\bibitem[{{Gaia Collaboration} {et~al.}(2018){Gaia Collaboration}, {Brown},
  {Vallenari}, {Prusti}, {de Bruijne}, {Babusiaux}, {Bailer-Jones}, {Biermann},
  {Evans}, {Eyer}, {Jansen}, {Jordi}, {Klioner}, {Lammers}, {Lindegren},
  {Luri}, {Mignard}, {Panem}, {Pourbaix}, {Randich}, {Sartoretti}, {Siddiqui},
  {Soubiran}, {van Leeuwen}, {Walton}, {Arenou}, {Bastian}, {Cropper},
  {Drimmel}, {Katz}, {Lattanzi}, {Bakker}, {Cacciari}, {Casta{\~n}eda},
  {Chaoul}, {Cheek}, {De Angeli}, {Fabricius}, {Guerra}, {Holl}, {Masana},
  {Messineo}, {Mowlavi}, {Nienartowicz}, {Panuzzo}, {Portell}, {Riello},
  {Seabroke}, {Tanga}, {Th{\'e}venin}, {Gracia-Abril}, {Comoretto},
  {Garcia-Reinaldos}, {Teyssier}, {Altmann}, {Andrae}, {Audard},
  {Bellas-Velidis}, {Benson}, {Berthier}, {Blomme}, {Burgess}, {Busso},
  {Carry}, {Cellino}, {Clementini}, {Clotet}, {Creevey}, {Davidson}, {De
  Ridder}, {Delchambre}, {Dell'Oro}, {Ducourant},
  {Fern{\'a}ndez-Hern{\'a}ndez}, {Fouesneau}, {Fr{\'e}mat}, {Galluccio},
  {Garc{\'\i}a-Torres}, {Gonz{\'a}lez-N{\'u}{\~n}ez}, {Gonz{\'a}lez-Vidal},
  {Gosset}, {Guy}, {Halbwachs}, {Hambly}, {Harrison}, {Hern{\'a}ndez},
  {Hestroffer}, {Hodgkin}, {Hutton}, {Jasniewicz}, {Jean-Antoine-Piccolo},
  {Jordan}, {Korn}, {Krone-Martins}, {Lanzafame}, {Lebzelter}, {L{\"o}ffler},
  {Manteiga}, {Marrese}, {Mart{\'\i}n-Fleitas}, {Moitinho}, {Mora}, {Muinonen},
  {Osinde}, {Pancino}, {Pauwels}, {Petit}, {Recio-Blanco}, {Richards},
  {Rimoldini}, {Robin}, {Sarro}, {Siopis}, {Smith}, {Sozzetti}, {S{\"u}veges},
  {Torra}, {van Reeven}, {Abbas}, {Abreu Aramburu}, {Accart}, {Aerts},
  {Altavilla}, {{\'A}lvarez}, {Alvarez}, {Alves}, {Anderson}, {Andrei},
  {Anglada Varela}, {Antiche}, {Antoja}, {Arcay}, {Astraatmadja}, {Bach},
  {Baker}, {Balaguer-N{\'u}{\~n}ez}, {Balm}, {Barache}, {Barata}, {Barbato},
  {Barblan}, {Barklem}, {Barrado}, {Barros}, {Barstow}, {Bartholom{\'e}
  Mu{\~n}oz}, {Bassilana}, {Becciani}, {Bellazzini}, {Berihuete}, {Bertone},
  {Bianchi}, {Bienaym{\'e}}, {Blanco-Cuaresma}, {Boch}, {Boeche}, {Bombrun},
  {Borrachero}, {Bossini}, {Bouquillon}, {Bourda}, {Bragaglia}, {Bramante},
  {Breddels}, {Bressan}, {Brouillet}, {Br{\"u}semeister}, {Brugaletta},
  {Bucciarelli}, {Burlacu}, {Busonero}, {Butkevich}, {Buzzi}, {Caffau},
  {Cancelliere}, {Cannizzaro}, {Cantat-Gaudin}, {Carballo}, {Carlucci},
  {Carrasco}, {Casamiquela}, {Castellani}, {Castro-Ginard}, {Charlot},
  {Chemin}, {Chiavassa}, {Cocozza}, {Costigan}, {Cowell}, {Crifo}, {Crosta},
  {Crowley}, {Cuypers}, {Dafonte}, {Damerdji}, {Dapergolas}, {David}, {David},
  {de Laverny}, {De Luise}, {De March}, {de Martino}, {de Souza}, {de Torres},
  {Debosscher}, {del Pozo}, {Delbo}, {Delgado}, {Delgado}, {Di Matteo},
  {Diakite}, {Diener}, {Distefano}, {Dolding}, {Drazinos}, {Dur{\'a}n},
  {Edvardsson}, {Enke}, {Eriksson}, {Esquej}, {Eynard Bontemps}, {Fabre},
  {Fabrizio}, {Faigler}, {Falc{\~a}o}, {Farr{\`a}s Casas}, {Federici},
  {Fedorets}, {Fernique}, {Figueras}, {Filippi}, {Findeisen}, {Fonti},
  {Fraile}, {Fraser}, {Fr{\'e}zouls}, {Gai}, {Galleti}, {Garabato},
  {Garc{\'\i}a-Sedano}, {Garofalo}, {Garralda}, {Gavel}, {Gavras}, {Gerssen},
  {Geyer}, {Giacobbe}, {Gilmore}, {Girona}, {Giuffrida}, {Glass}, {Gomes},
  {Granvik}, {Gueguen}, {Guerrier}, {Guiraud}, {Guti{\'e}rrez-S{\'a}nchez},
  {Haigron}, {Hatzidimitriou}, {Hauser}, {Haywood}, {Heiter}, {Helmi}, {Heu},
  {Hilger}, {Hobbs}, {Hofmann}, {Holland}, {Huckle}, {Hypki}, {Icardi},
  {Jan{\ss}en}, {Jevardat de Fombelle}, {Jonker}, {Juh{\'a}sz}, {Julbe},
  {Karampelas}, {Kewley}, {Klar}, {Kochoska}, {Kohley}, {Kolenberg},
  {Kontizas}, {Kontizas}, {Koposov}, {Kordopatis}, {Kostrzewa-Rutkowska},
  {Koubsky}, {Lambert}, {Lanza}, {Lasne}, {Lavigne}, {Le Fustec}, {Le
  Poncin-Lafitte}, {Lebreton}, {Leccia}, {Leclerc}, {Lecoeur-Taibi},
  {Lenhardt}, {Leroux}, {Liao}, {Licata}, {Lindstr{\o}m}, {Lister}, {Livanou},
  {Lobel}, {L{\'o}pez}, {Managau}, {Mann}, {Mantelet}, {Marchal}, {Marchant},
  {Marconi}, {Marinoni}, {Marschalk{\'o}}, {Marshall}, {Martino}, {Marton},
  {Mary}, {Massari}, {Matijevi{\v{c}}}, {Mazeh}, {McMillan}, {Messina},
  {Michalik}, {Millar}, {Molina}, {Molinaro}, {Moln{\'a}r}, {Montegriffo},
  {Mor}, {Morbidelli}, {Morel}, {Morris}, {Mulone}, {Muraveva}, {Musella},
  {Nelemans}, {Nicastro}, {Noval}, {O'Mullane}, {Ord{\'e}novic},
  {Ord{\'o}{\~n}ez-Blanco}, {Osborne}, {Pagani}, {Pagano}, {Pailler},
  {Palacin}, {Palaversa}, {Panahi}, {Pawlak}, {Piersimoni}, {Pineau}, {Plachy},
  {Plum}, {Poggio}, {Poujoulet}, {Pr{\v{s}}a}, {Pulone}, {Racero}, {Ragaini},
  {Rambaux}, {Ramos-Lerate}, {Regibo}, {Reyl{\'e}}, {Riclet}, {Ripepi}, {Riva},
  {Rivard}, {Rixon}, {Roegiers}, {Roelens}, {Romero-G{\'o}mez}, {Rowell},
  {Royer}, {Ruiz-Dern}, {Sadowski}, {Sagrist{\`a} Sell{\'e}s}, {Sahlmann},
  {Salgado}, {Salguero}, {Sanna}, {Santana-Ros}, {Sarasso}, {Savietto},
  {Schultheis}, {Sciacca}, {Segol}, {Segovia}, {S{\'e}gransan}, {Shih},
  {Siltala}, {Silva}, {Smart}, {Smith}, {Solano}, {Solitro}, {Sordo}, {Soria
  Nieto}, {Souchay}, {Spagna}, {Spoto}, {Stampa}, {Steele},
  {Steidelm{\"u}ller}, {Stephenson}, {Stoev}, {Suess}, {Surdej}, {Szabados},
  {Szegedi-Elek}, {Tapiador}, {Taris}, {Tauran}, {Taylor}, {Teixeira},
  {Terrett}, {Teyssand ier}, {Thuillot}, {Titarenko}, {Torra Clotet}, {Turon},
  {Ulla}, {Utrilla}, {Uzzi}, {Vaillant}, {Valentini}, {Valette}, {van Elteren},
  {Van Hemelryck}, {van Leeuwen}, {Vaschetto}, {Vecchiato}, {Veljanoski},
  {Viala}, {Vicente}, {Vogt}, {von Essen}, {Voss}, {Votruba}, {Voutsinas},
  {Walmsley}, {Weiler}, {Wertz}, {Wevers}, {Wyrzykowski}, {Yoldas},
  {{\v{Z}}erjal}, {Ziaeepour}, {Zorec}, {Zschocke}, {Zucker}, {Zurbach}, \&
  {Zwitter}}]{GAIADR2}
{Gaia Collaboration}, {Brown}, A.~G.~A., {Vallenari}, A., {et~al.} 2018, \aap,
  616, A1, \dodoi{10.1051/0004-6361/201833051}

\bibitem[{Geach {et~al.}(2011)Geach, Smail, Moran, MacArthur, Lagos, \&
  Edge}]{Geach_2011}
Geach, J.~E., Smail, I., Moran, S.~M., {et~al.} 2011, The Astrophysical
  Journal, 730, L19, \dodoi{10.1088/2041-8205/730/2/l19}

\bibitem[{Genzel {et~al.}(2012)Genzel, Tacconi, Combes, Bolatto, Neri,
  Sternberg, Cooper, Bouché, Bournaud, Burkert, \& et~al.}]{Genzel_2012}
Genzel, R., Tacconi, L.~J., Combes, F., {et~al.} 2012, The Astrophysical
  Journal, 746, 69, \dodoi{10.1088/0004-637x/746/1/69}

\bibitem[{{Genzel} {et~al.}(2015){Genzel}, {Tacconi}, {Lutz}, {Saintonge},
  {Berta}, {Magnelli}, {Combes}, {Garc{\'\i}a-Burillo}, {Neri}, {Bolatto},
  {Contini}, {Lilly}, {Boissier}, {Boone}, {Bouch{\'e}}, {Bournaud}, {Burkert},
  {Carollo}, {Colina}, {Cooper}, {Cox}, {Feruglio}, {F{\"o}rster Schreiber},
  {Freundlich}, {Gracia-Carpio}, {Juneau}, {Kovac}, {Lippa}, {Naab}, {Salome},
  {Renzini}, {Sternberg}, {Walter}, {Weiner}, {Weiss}, \& {Wuyts}}]{Genzel2015}
{Genzel}, R., {Tacconi}, L.~J., {Lutz}, D., {et~al.} 2015, \apj, 800, 20,
  \dodoi{10.1088/0004-637X/800/1/20}

\bibitem[{{Ginsburg} {et~al.}(2019){Ginsburg}, {Koch}, {Robitaille},
  {Beaumont}, {adamginsburg}, {ZuHone}, {Sipocz}, {Patra}, {Jones}, {Lim},
  {Rosolowsky}, {Stern}, {Earl}, {de Val-Borro}, {jrobbfed}, {shuokong},
  {Kepley}, {Sokolov}, {Badger}, {Maret}, {Garrido}, {Booker}, \&
  {Tollerud}}]{spectral-cube}
{Ginsburg}, A., {Koch}, E., {Robitaille}, T., {et~al.} 2019,
  {radio-astro-tools/spectral-cube: v0.4.4}, v0.4.4,  Zenodo,
  \dodoi{10.5281/zenodo.2573901}

\bibitem[{Gomez {et~al.}(2003)Gomez, Nichol, Miller, Balogh, Goto, Zabludoff,
  Romer, Bernardi, Sheth, Hopkins, \& et~al.}]{Gomez_2003}
Gomez, P.~L., Nichol, R.~C., Miller, C.~J., {et~al.} 2003, The Astrophysical
  Journal, 584, 210–227, \dodoi{10.1086/345593}

\bibitem[{{G{\'o}mez-Guijarro} {et~al.}(2019){G{\'o}mez-Guijarro}, {Riechers},
  {Pavesi}, {Magdis}, {Leung}, {Valentino}, {Toft}, {Aravena}, {Chapman},
  {Clements}, {Dannerbauer}, {Oliver}, {P{\'e}rez-Fournon}, \&
  {Valtchanov}}]{Gomez-Guijarro_2019}
{G{\'o}mez-Guijarro}, C., {Riechers}, D.~A., {Pavesi}, R., {et~al.} 2019, \apj,
  872, 117, \dodoi{10.3847/1538-4357/ab002a}

\bibitem[{Hayashi {et~al.}(2018)Hayashi, Tadaki, Kodama, Kohno, Yamaguchi,
  Hatsukade, Koyama, Shimakawa, Tamura, \& Suzuki}]{Hayashi2018}
Hayashi, M., Tadaki, K.-i., Kodama, T., {et~al.} 2018, The Astrophysical
  Journal, 856, 118, \dodoi{10.3847/1538-4357/aab3e7}

\bibitem[{Herrero-Illana {et~al.}(2019)Herrero-Illana, Privon, Evans,
  Díaz-Santos, Pérez-Torres, U, Alberdi, Iwasawa, Armus, Aalto, \&
  et~al.}]{Herrero_Illana_2019}
Herrero-Illana, R., Privon, G.~C., Evans, A.~S., {et~al.} 2019, Astronomy {\&}
  Astrophysics, 628, A71, \dodoi{10.1051/0004-6361/201834088}

\bibitem[{Hodge {et~al.}(2013)Hodge, Carilli, Walter, Daddi, \&
  Riechers}]{Hodge_2013}
Hodge, J.~A., Carilli, C.~L., Walter, F., Daddi, E., \& Riechers, D. 2013, The
  Astrophysical Journal, 776, 22, \dodoi{10.1088/0004-637x/776/1/22}

\bibitem[{Hodge \& da~Cunha(2020)}]{Hodge2020}
Hodge, J.~A., \& da~Cunha, E. 2020.
\newblock \doarXiv{2004.00934}

\bibitem[{{H{\"o}gbom}(1974)}]{Hogbom_1974}
{H{\"o}gbom}, J.~A. 1974, \aaps, 15, 417

\bibitem[{Hunter(2007)}]{Hunter_2007}
Hunter, J.~D. 2007, Matplotlib: A 2D graphics environment,  IEEE COMPUTER SOC,
  \dodoi{10.1109/MCSE.2007.55}

\bibitem[{Ivison {et~al.}(2011)Ivison, Papadopoulos, Smail, Greve, Thomson,
  Xilouris, \& Chapman}]{Ivison2011}
Ivison, R.~J., Papadopoulos, P.~P., Smail, I., {et~al.} 2011, Monthly Notices
  of the Royal Astronomical Society, 412, 1913,
  \dodoi{10.1111/j.1365-2966.2010.18028.x}

\bibitem[{Jauzac {et~al.}(2020)Jauzac, Klein, Kneib, Richard, Rexroth,
  Sch{\"{a}}fer, \& Verdier}]{Jauzac2020}
Jauzac, M., Klein, B., Kneib, J.-P., {et~al.} 2020, 22, 1.
\newblock \doarXiv{2006.10700}

\bibitem[{Kennicutt(1998)}]{Kennicutt_1998}
Kennicutt, R.~C. 1998, Annual Review of Astronomy and Astrophysics, 36,
  189–231, \dodoi{10.1146/annurev.astro.36.1.189}

\bibitem[{Kroupa(2001)}]{Kroupa_2001}
Kroupa, P. 2001, Monthly Notices of the Royal Astronomical Society, 322,
  231–246, \dodoi{10.1046/j.1365-8711.2001.04022.x}

\bibitem[{Lee {et~al.}(2017)Lee, Tanaka, Kawabe, Kohno, Kodama, Kajisawa, Yun,
  Nakanishi, Iono, Tamura, \& et~al.}]{Lee_2017}
Lee, M.~M., Tanaka, I., Kawabe, R., {et~al.} 2017, The Astrophysical Journal,
  842, 55, \dodoi{10.3847/1538-4357/aa74c2}

\bibitem[{Levenberg(1944)}]{Levenberg1944}
Levenberg, K. 1944, Quarterly of Applied Mathematics, 2, 164,
  \dodoi{10.1090/qam/10666}

\bibitem[{{Lotz} {et~al.}(2013){Lotz}, {Papovich}, {Faber}, {Ferguson},
  {Grogin}, {Guo}, {Kocevski}, {Koekemoer}, {Lee}, {McIntosh}, {Momcheva},
  {Rudnick}, {Saintonge}, {Tran}, {van der Wel}, \& {Willmer}}]{Lotz2013}
{Lotz}, J.~M., {Papovich}, C., {Faber}, S.~M., {et~al.} 2013, \apj, 773, 154,
  \dodoi{10.1088/0004-637X/773/2/154}

\bibitem[{MacKenzie {et~al.}(2014)MacKenzie, Scott, Smail, Chapin, Chapman,
  Conley, Cooray, Dunlop, Farrah, Fich, Gibb, Holland, Ivison, Jenness, Kneib,
  Marsden, Richard, Robson, Valtchanov, \& Wardlow}]{MacKenzie2014}
MacKenzie, T.~P., Scott, D., Smail, I., {et~al.} 2014, Monthly Notices of the
  Royal Astronomical Society, 445, 201, \dodoi{10.1093/mnras/stu1623}

\bibitem[{Madau \& Dickinson(2014)}]{Madau_2014}
Madau, P., \& Dickinson, M. 2014, Annual Review of Astronomy and Astrophysics,
  52, 415–486, \dodoi{10.1146/annurev-astro-081811-125615}

\bibitem[{{Magdis} {et~al.}(2012){Magdis}, {Daddi}, {Sargent}, {Elbaz},
  {Gobat}, {Dannerbauer}, {Feruglio}, {Tan}, {Rigopoulou}, {Charmandaris},
  {Dickinson}, {Reddy}, \& {Aussel}}]{Magdis_2012}
{Magdis}, G.~E., {Daddi}, E., {Sargent}, M., {et~al.} 2012, \apjl, 758, L9,
  \dodoi{10.1088/2041-8205/758/1/L9}

\bibitem[{Magdis {et~al.}(2014)Magdis, Rigopoulou, Hopwood, Huang, Farrah,
  Pearson, Alonso-Herrero, Bock, Clements, Cooray, \& et~al.}]{Magdis_2014}
Magdis, G.~E., Rigopoulou, D., Hopwood, R., {et~al.} 2014, The Astrophysical
  Journal, 796, 63, \dodoi{10.1088/0004-637x/796/1/63}

\bibitem[{Man \& Belli(2018)}]{Man_2018}
Man, A., \& Belli, S. 2018, Nature Astronomy, 2, 695–697,
  \dodoi{10.1038/s41550-018-0558-1}

\bibitem[{{McMullin} {et~al.}(2007){McMullin}, {Waters}, {Schiebel}, {Young},
  \& {Golap}}]{McMullin2007}
{McMullin}, J.~P., {Waters}, B., {Schiebel}, D., {Young}, W., \& {Golap}, K.
  2007, in Astronomical Society of the Pacific Conference Series, Vol. 376,
  Astronomical Data Analysis Software and Systems XVI, ed. R.~A. {Shaw},
  F.~{Hill}, \& D.~J. {Bell}, 127

\bibitem[{Miller {et~al.}(2018)Miller, Chapman, Aravena, Ashby, Hayward,
  Vieira, Wei{\ss}, Babul, B{\'{e}}thermin, Bradford, Brodwin, Carlstrom, Chen,
  Cunningham, {De Breuck}, Gonzalez, Greve, Harnett, Hezaveh, Lacaille, Litke,
  Ma, Malkan, Marrone, Morningstar, Murphy, Narayanan, Pass, Perry, Phadke,
  Rennehan, Rotermund, Simpson, Spilker, Sreevani, Stark, Strandet, \&
  Strom}]{Miller2018}
Miller, T.~B., Chapman, S.~C., Aravena, M., {et~al.} 2018, in Nature,
  \dodoi{10.1038/s41586-018-0025-2}

\bibitem[{{Mo} {et~al.}(2010){Mo}, {van den Bosch}, \& {White}}]{Mo_2010}
{Mo}, H., {van den Bosch}, F.~C., \& {White}, S. 2010, {Galaxy Formation and
  Evolution}

\bibitem[{Muldrew {et~al.}(2015)Muldrew, Hatch, \& Cooke}]{Muldrew_2015}
Muldrew, S.~I., Hatch, N.~A., \& Cooke, E.~A. 2015, Monthly Notices of the
  Royal Astronomical Society, 452, 2528–2539, \dodoi{10.1093/mnras/stv1449}

\bibitem[{{Murphy} {et~al.}(2011){Murphy}, {Condon}, {Schinnerer}, {Kennicutt},
  {Calzetti}, {Armus}, {Helou}, {Turner}, {Aniano}, {Beir{\~a}o}, {Bolatto},
  {Brandl}, {Croxall}, {Dale}, {Donovan Meyer}, {Draine}, {Engelbracht},
  {Hunt}, {Hao}, {Koda}, {Roussel}, {Skibba}, \& {Smith}}]{Murphy2011}
{Murphy}, E.~J., {Condon}, J.~J., {Schinnerer}, E., {et~al.} 2011, \apj, 737,
  67, \dodoi{10.1088/0004-637X/737/2/67}

\bibitem[{{Narayanan} \& {Krumholz}(2014)}]{Narayanan2014}
{Narayanan}, D., \& {Krumholz}, M.~R. 2014, \mnras, 442, 1411,
  \dodoi{10.1093/mnras/stu834}

\bibitem[{Narayanan {et~al.}(2012)Narayanan, Krumholz, Ostriker, \&
  Hernquist}]{Narayanan_2012}
Narayanan, D., Krumholz, M.~R., Ostriker, E.~C., \& Hernquist, L. 2012, Monthly
  Notices of the Royal Astronomical Society, 421, 3127–3146,
  \dodoi{10.1111/j.1365-2966.2012.20536.x}

\bibitem[{Noble {et~al.}(2017)Noble, McDonald, Muzzin, Nantais, Rudnick, van
  Kampen, Webb, Wilson, Yee, Boone, Cooper, DeGroot, Delahaye, Demarco, Foltz,
  Hayden, Lidman, Manilla-Robles, \& Perlmutter}]{Noble2017}
Noble, A.~G., McDonald, M., Muzzin, A., {et~al.} 2017, The Astrophysical
  Journal, 842, L21, \dodoi{10.3847/2041-8213/aa77f3}

\bibitem[{Noble {et~al.}(2019)Noble, Muzzin, McDonald, Rudnick, Matharu,
  Cooper, Demarco, Lidman, Nantais, van Kampen, Webb, Wilson, \&
  Yee}]{Noble2019}
Noble, A.~G., Muzzin, A., McDonald, M., {et~al.} 2019, The Astrophysical
  Journal, 870, 56, \dodoi{10.3847/1538-4357/aaf1c6}

\bibitem[{{Oteo} {et~al.}(2018){Oteo}, {Ivison}, {Dunne}, {Manilla-Robles},
  {Maddox}, {Lewis}, {de Zotti}, {Bremer}, {Clements}, {Cooray}, {Dannerbauer},
  {Eales}, {Greenslade}, {Omont}, {Perez{\textendash}Fourn{\'o}n}, {Riechers},
  {Scott}, {van der Werf}, {Weiss}, \& {Zhang}}]{Oteo2018}
{Oteo}, I., {Ivison}, R.~J., {Dunne}, L., {et~al.} 2018, \apj, 856, 72,
  \dodoi{10.3847/1538-4357/aaa1f1}

\bibitem[{Papadopoulos {et~al.}(2012)Papadopoulos, van~der Werf, Xilouris,
  Isaak, Gao, \& M{\"{u}}hle}]{Papadopoulos2012}
Papadopoulos, P.~P., van~der Werf, P.~P., Xilouris, E.~M., {et~al.} 2012,
  Monthly Notices of the Royal Astronomical Society, 426, 2601,
  \dodoi{10.1111/j.1365-2966.2012.21001.x}

\bibitem[{Popesso {et~al.}(2015{\natexlab{a}})Popesso, Biviano, Finoguenov,
  Wilman, Salvato, Magnelli, Gruppioni, Pozzi, Rodighiero, Ziparo, \&
  et~al.}]{Popesso_2015a}
Popesso, P., Biviano, A., Finoguenov, A., {et~al.} 2015{\natexlab{a}},
  Astronomy {\&} Astrophysics, 574, A105, \dodoi{10.1051/0004-6361/201424711}

\bibitem[{Popesso {et~al.}(2015{\natexlab{b}})Popesso, Biviano, Finoguenov,
  Wilman, Salvato, Magnelli, Gruppioni, Pozzi, Rodighiero, Ziparo, \&
  et~al.}]{Popesso_2015b}
---. 2015{\natexlab{b}}, Astronomy {\&} Astrophysics, 579, A132,
  \dodoi{10.1051/0004-6361/201424715}

\bibitem[{Price-Whelan {et~al.}(2018)Price-Whelan, Sipőcz, Günther, Lim,
  Crawford, Conseil, Shupe, Craig, Dencheva, \& et~al.}]{Astropy_2018}
Price-Whelan, A.~M., Sipőcz, B.~M., Günther, H.~M., {et~al.} 2018, The
  Astropy Project: Building an Open-science Project and Status of the v2.0 Core
  Package,  American Astronomical Society, \dodoi{10.3847/1538-3881/aabc4f}

\bibitem[{Riechers {et~al.}(2010{\natexlab{a}})Riechers, Carilli, Walter, \&
  Momjian}]{Riechers_2010a}
Riechers, D.~A., Carilli, C.~L., Walter, F., \& Momjian, E. 2010{\natexlab{a}},
  The Astrophysical Journal, 724, L153–L157,
  \dodoi{10.1088/2041-8205/724/2/l153}

\bibitem[{{Riechers} {et~al.}(2007){Riechers}, {Walter}, {Carilli}, {Knudsen},
  {Lo}, {Benford}, {Staguhn}, {Hunter}, {Bertoldi}, {Henkel}, {Menten},
  {Weiss}, {Yun}, \& {Scoville}}]{Riechers_2007}
{Riechers}, D.~A., {Walter}, F., {Carilli}, C.~L., {et~al.} 2007, in
  Astronomical Society of the Pacific Conference Series, Vol. 375, From
  Z-Machines to ALMA: (Sub)Millimeter Spectroscopy of Galaxies, ed. A.~J.
  {Baker}, J.~{Glenn}, A.~I. {Harris}, J.~G. {Mangum}, \& M.~S. {Yun}, 148

\bibitem[{Riechers {et~al.}(2010{\natexlab{b}})Riechers, Capak, Carilli, Cox,
  Neri, Scoville, Schinnerer, Bertoldi, \& Yan}]{Riechers_2010b}
Riechers, D.~A., Capak, P.~L., Carilli, C.~L., {et~al.} 2010{\natexlab{b}}, The
  Astrophysical Journal, 720, L131–L136, \dodoi{10.1088/2041-8205/720/2/l131}

\bibitem[{Robitaille {et~al.}(2013)Robitaille, Tollerud, Greenfield,
  Droettboom, Bray, Aldcroft, Davis, Ginsburg, Price-Whelan, \&
  et~al.}]{Astropy_2013}
Robitaille, T.~P., Tollerud, E.~J., Greenfield, P., {et~al.} 2013, Astropy: A
  community Python package for astronomy,  EDP Sciences,
  \dodoi{10.1051/0004-6361/201322068}

\bibitem[{Rudnick {et~al.}(2017)Rudnick, Hodge, Walter, Momcheva, Tran,
  Papovich, da~Cunha, Decarli, Saintonge, Willmer, \& et~al.}]{Rudnick_2017}
Rudnick, G., Hodge, J., Walter, F., {et~al.} 2017, The Astrophysical Journal,
  849, 27, \dodoi{10.3847/1538-4357/aa87b2}

\bibitem[{Saintonge {et~al.}(2013)Saintonge, Lutz, Genzel, Magnelli, Nordon,
  Tacconi, Baker, Bandara, Berta, Förster~Schreiber, \&
  et~al.}]{Saintonge_2013}
Saintonge, A., Lutz, D., Genzel, R., {et~al.} 2013, The Astrophysical Journal,
  778, 2, \dodoi{10.1088/0004-637x/778/1/2}

\bibitem[{{Schinnerer} {et~al.}(2010){Schinnerer}, {Sargent}, {Bondi},
  {Smol{\v{c}}i{\'c}}, {Datta}, {Carilli}, {Bertoldi}, {Blain}, {Ciliegi},
  {Koekemoer}, \& {Scoville}}]{Schinnerer2010}
{Schinnerer}, E., {Sargent}, M.~T., {Bondi}, M., {et~al.} 2010, \apjs, 188,
  384, \dodoi{10.1088/0067-0049/188/2/384}

\bibitem[{{Schmidt}(1959)}]{Schmidt_1959}
{Schmidt}, M. 1959, \apj, 129, 243, \dodoi{10.1086/146614}

\bibitem[{Shapley(2011)}]{Shapley_2011}
Shapley, A.~E. 2011, Annual Review of Astronomy and Astrophysics, 49,
  525–580, \dodoi{10.1146/annurev-astro-081710-102542}

\bibitem[{Solomon {et~al.}(1992)Solomon, Downes, \& Radford}]{Solomon1992}
Solomon, P.~M., Downes, D., \& Radford, S. J.~E. 1992, The Astrophysical
  Journal, 398, L29, \dodoi{10.1086/186569}

\bibitem[{Spilker {et~al.}(2015)Spilker, Aravena, Marrone, Béthermin,
  Bothwell, Carlstrom, Chapman, Collier, Breuck, Fassnacht, \&
  et~al.}]{Spilker_2015}
Spilker, J.~S., Aravena, M., Marrone, D.~P., {et~al.} 2015, The Astrophysical
  Journal, 811, 124, \dodoi{10.1088/0004-637x/811/2/124}

\bibitem[{Spingola {et~al.}(2020)Spingola, McKean, Vegetti, Powell, Auger,
  Koopmans, Fassnacht, Lagattuta, Rizzo, Stacey, \& et~al.}]{Spingola_2020}
Spingola, C., McKean, J.~P., Vegetti, S., {et~al.} 2020, Monthly Notices of the
  Royal Astronomical Society, 495, 2387–2407, \dodoi{10.1093/mnras/staa1342}

\bibitem[{Stach {et~al.}(2017)Stach, Swinbank, Smail, Hilton, Simpson, \&
  Cooke}]{Stach_2017}
Stach, S.~M., Swinbank, A.~M., Smail, I., {et~al.} 2017, The Astrophysical
  Journal, 849, 154, \dodoi{10.3847/1538-4357/aa93f6}

\bibitem[{Strazzullo {et~al.}(2019)Strazzullo, Pannella, Mohr, Saro, Ashby,
  Bayliss, Bocquet, Bulbul, Khullar, Mantz, \& et~al.}]{Strazzullo_2019}
Strazzullo, V., Pannella, M., Mohr, J.~J., {et~al.} 2019, Astronomy {\&}
  Astrophysics, 622, A117, \dodoi{10.1051/0004-6361/201833944}

\bibitem[{Tacconi {et~al.}(2020)Tacconi, Genzel, \& Sternberg}]{Tacconi2020}
Tacconi, L.~J., Genzel, R., \& Sternberg, A. 2020.
\newblock \doarXiv{2003.06245}

\bibitem[{Tacconi {et~al.}(2013)Tacconi, Neri, Genzel, Combes, Bolatto, Cooper,
  Wuyts, Bournaud, Burkert, Comerford, Cox, Davis, {F{\"{o}}rster Schreiber},
  Garc{\'{i}}a-Burillo, Gracia-Carpio, Lutz, Naab, Newman, Omont, Saintonge,
  {Shapiro Griffin}, Shapley, Sternberg, \& Weiner}]{Tacconi2013}
Tacconi, L.~J., Neri, R., Genzel, R., {et~al.} 2013, The Astrophysical Journal,
  768, 74, \dodoi{10.1088/0004-637X/768/1/74}

\bibitem[{Tacconi {et~al.}(2018)Tacconi, Genzel, Saintonge, Combes,
  García-Burillo, Neri, Bolatto, Contini, Schreiber, Lilly, \&
  et~al.}]{Tacconi_2018}
Tacconi, L.~J., Genzel, R., Saintonge, A., {et~al.} 2018, The Astrophysical
  Journal, 853, 179, \dodoi{10.3847/1538-4357/aaa4b4}

\bibitem[{Tadaki {et~al.}(2018)Tadaki, Iono, Yun, Aretxaga, Hatsukade, Hughes,
  Ikarashi, Izumi, Kawabe, Kohno, Lee, Matsuda, Nakanishi, Saito, Tamura, Ueda,
  Umehata, Wilson, Michiyama, Ando, \& Kamieneski}]{Tadaki2018}
Tadaki, K., Iono, D., Yun, M.~S., {et~al.} 2018, Nature, 560, 613,
  \dodoi{10.1038/s41586-018-0443-1}

\bibitem[{Tadaki {et~al.}(2019)Tadaki, Kodama, Hayashi, Shimakawa, Koyama, Lee,
  Tanaka, Hatsukade, Iono, Kohno, Matsuda, Suzuki, Tamura, Toshikawa, \&
  Umehata}]{Tadaki2019}
Tadaki, K.-i., Kodama, T., Hayashi, M., {et~al.} 2019, Publications of the
  Astronomical Society of Japan, 71, 1, \dodoi{10.1093/pasj/psz005}

\bibitem[{Tan {et~al.}(2013)Tan, Daddi, Sargent, Magdis, Hodge, Béthermin,
  Bournaud, Carilli, Dannerbauer, Dickinson, \& et~al.}]{Tan_2013}
Tan, Q., Daddi, E., Sargent, M., {et~al.} 2013, The Astrophysical Journal, 776,
  L24, \dodoi{10.1088/2041-8205/776/2/l24}

\bibitem[{Tran {et~al.}(2010)Tran, Papovich, Saintonge, Brodwin, Dunlop,
  Farrah, Finkelstein, Finkelstein, Lotz, McLure, \& et~al.}]{Tran_2010}
Tran, K.-V.~H., Papovich, C., Saintonge, A., {et~al.} 2010, The Astrophysical
  Journal, 719, L126–L129, \dodoi{10.1088/2041-8205/719/2/l126}

\bibitem[{van~der Walt {et~al.}(2011)van~der Walt, Colbert, \&
  Varoquaux}]{van_der_Walt_2011}
van~der Walt, S., Colbert, S.~C., \& Varoquaux, G. 2011, The NumPy Array: A
  Structure for Efficient Numerical Computation,  Institute of Electrical and
  Electronics Engineers (IEEE), \dodoi{10.1109/mcse.2011.37}

\bibitem[{Vegetti \& Koopmans(2009)}]{Vegetti_2009}
Vegetti, S., \& Koopmans, L. V.~E. 2009, Monthly Notices of the Royal
  Astronomical Society, 392, 945–963,
  \dodoi{10.1111/j.1365-2966.2008.14005.x}

\bibitem[{{Virtanen} {et~al.}(2020){Virtanen}, {Gommers}, {Oliphant},
  {Haberland}, {Reddy}, {Cournapeau}, {Burovski}, {Peterson}, {Weckesser},
  {Bright}, {van der Walt}, {Brett}, {Wilson}, {Jarrod Millman}, {Mayorov},
  {Nelson}, {Jones}, {Kern}, {Larson}, {Carey}, {Polat}, {Feng}, {Moore}, {Vand
  erPlas}, {Laxalde}, {Perktold}, {Cimrman}, {Henriksen}, {Quintero}, {Harris},
  {Archibald}, {Ribeiro}, {Pedregosa}, {van Mulbregt}, \&
  {Contributors}}]{SciPy_2020}
{Virtanen}, P., {Gommers}, R., {Oliphant}, T.~E., {et~al.} 2020, Nature
  Methods, 17, 261, \dodoi{https://doi.org/10.1038/s41592-019-0686-2}

\bibitem[{Walter {et~al.}(2004)Walter, Carilli, Bertoldi, Menten, Cox, Lo, Fan,
  \& Strauss}]{Walter_2004}
Walter, F., Carilli, C., Bertoldi, F., {et~al.} 2004, The Astrophysical
  Journal, 615, L17–L20, \dodoi{10.1086/426017}

\bibitem[{Wang {et~al.}(2018)Wang, Elbaz, Daddi, Liu, Kodama, Tanaka,
  Schreiber, Zanella, Valentino, Sargent, Kohno, Xiao, Pannella, Ciesla, Gobat,
  \& Koyama}]{Wang2018}
Wang, T., Elbaz, D., Daddi, E., {et~al.} 2018, The Astrophysical Journal, 867,
  L29, \dodoi{10.3847/2041-8213/aaeb2c}

\bibitem[{Waskom \& the seaborn~development team(2020)}]{waskom2020seaborn}
Waskom, M., \& the seaborn~development team. 2020, mwaskom/seaborn, latest,
  Zenodo, \dodoi{10.5281/zenodo.592845}

\bibitem[{Zavala {et~al.}(2019)Zavala, Casey, Scoville, Champagne, Chiang,
  Dannerbauer, Drew, Fu, Spilker, Spitler, Tran, Treister, \&
  Toft}]{Zavala2019}
Zavala, J.~A., Casey, C.~M., Scoville, N., {et~al.} 2019, The Astrophysical
  Journal, 887, 183, \dodoi{10.3847/1538-4357/ab5302}

\bibitem[{{Zhang} {et~al.}(2016){Zhang}, {Papadopoulos}, {Ivison}, {Galametz},
  {Smith}, \& {Xilouris}}]{Zhang_2016}
{Zhang}, Z.-Y., {Papadopoulos}, P.~P., {Ivison}, R.~J., {et~al.} 2016, Royal
  Society Open Science, 3, 160025, \dodoi{10.1098/rsos.160025}

\bibitem[{Zitrin {et~al.}(2010)Zitrin, Broadhurst, Barkana, Rephaeli, \&
  Benítez}]{Zitrin_2010}
Zitrin, A., Broadhurst, T., Barkana, R., Rephaeli, Y., \& Benítez, N. 2010,
  Monthly Notices of the Royal Astronomical Society, no–no,
  \dodoi{10.1111/j.1365-2966.2010.17574.x}

\bibitem[{{Zwicky}(1956)}]{Zwicky1956}
{Zwicky}, F. 1956, Ergebnisse der exakten Naturwissenschaften, 29, 344

\end{thebibliography}

\appendix

\Needspace{4\baselineskip}
\section{Compilation}\label{sec:appendix-compilation}
\begin{deluxetable*}{cccCCCCC}[!ht]
  \tabletypesize{\footnotesize}
  \tablecaption{Comparison sample, ordered by maximum redshift. \label{table:compilation}}
  \tablehead{
    \colhead{Reference} & \colhead{Overdense} & \colhead{Starburst} & \colhead{Objects} & \colhead{Redshift Range} & \colhead{Gas Mass Range} & \colhead{SFR Range} & \colhead{Depletion Time Range}
    \\
    \colhead{} & \colhead{} & \colhead{} & \colhead{} & \colhead{} & \colhead{$\times 10^{10}~{\rm M_\odot}$} & \colhead{${\rm M_\odot\,yr^{-1}}$} & \colhead{Gyr}
  }
  \startdata
  \cite{Cicone_2017}                                      & no  & no  & 97 & 0.011-0.029 & 0.005-1.046 & 0-13     & 0.004-3.238 \\
  \cite{Herrero_Illana_2019}\tablenotemark{a}             & no  & yes & 53 & 0.011-0.088 & 0.090-3.017 & 20-450   & 0.016-0.183 \\
  \cite{Papadopoulos2012}\tablenotemark{a}                & no  & yes & 70 & 0.003-0.100 & 0.005-1.392 & 15-286   & 0.007-1.684 \\
  \cite{Cybulski_2016}                                    & yes & no  & 8  & 0.187-0.208 & 0.92-1.84   & 3-44     & 0.412-4.325 \\
  \cite{Geach_2011}                                       & yes & no  & 5  & 0.380-0.396 & 1.38-5.06   & 37-66    & 0.371-0.806 \\
  \cite{Magdis_2014}\tablenotemark{a}                     & no  & yes & 9  & 0.216-0.436 & 0.52-1.56   & 71-898   & 0.014-0.107 \\
  \cite{Combes_2013}\tablenotemark{a}                     & no  & yes & 39 & 0.607-0.996 & 0.08-3.36   & 201-5535 & 0.001-0.050 \\
  \cite{Hayashi2018}                                      & yes & no  & 17 & 1.445-1.471 & 2.7-10.7    & 3-154    & 0.443-10.030 \\
  \cite{Stach_2017}                                       & yes & yes & 6  & 1.450-1.472 & 1.0-2.4     & 43-149   & 0.067-0.414 \\
  \cite{Daddi_2010, Daddi_2015}                           & no  & no  & 6  & 1.414-1.600 & 4.2-12.0    & 66-425   & 0.282-0.703 \\
  \cite{Rudnick_2017}                                     & yes & no  & 2  & 1.624-1.629 & 3.31-11.22  & 12-165   & 0.678-2,595 \\
  \cite{Noble2017, Noble2019}                             & yes & no  & 17 & 1.596-1.635 & 1.7-25.5    & 3-230    & 0.248-31.970 \\
  \cite{Dannerbauer_2017}                                 & yes & no  & 1  & 2.148       & 18.0        & 372      & 0.484 \\
  \cite{Tacconi2013}                                      & no  & no  & 53 & 1.002-2.434 & 1.1-35.0    & 18-670   & 0.146-3.436 \\
  \cite{Ivison2011}\tablenotemark{a}                      & no  & yes & 5  & 2.201-2.487 & 2.5-9.8     & 568-1430 & 0.029-0.130 \\
  \cite{Lee_2017}                                         & yes & no  & 7  & 2.478-2.487 & 3.14-18.52  & 69-440   & 0.079-3.699 \\
  \cite{Zavala2019}                                       & yes & no  & 67 & 2.085-2.513 & 1.5-37.3    & 3-670    & 0.022-5.955 \\
  \cite{Wang2018}                                         & yes & no  & 14 & 2.494-2.515 & 1.4-55.0    & 14-796   & 0.200-6.170 \\
  \cite{Tadaki2019}                                       & yes & no  & 16 & 2.144-2.529 & 5.75-34.67  & 38-496   & 0.211-6.521 \\
  \cite{Gomez-Guijarro_2019}                              & yes & yes & 11 & 2.171-2.602 & 6.61-109.65 & 106-1808 & 0.236-1.525 \\
  \cite{Riechers_2010b}\tablenotemark{b}                  & no  & no  & 2  & 2.730-3.070 & 0.046-0.093 & 27-149   & 0.006-0.017 \\
  \cite{Coppin_2007}\tablenotemark{b}                     & no  & no  & 1  & 3.074       & 0.24        & 25       & 0.095 \\
  \cite{Saintonge_2013}\tablenotemark{b}                  & no  & no  & 10 & 1.411-3.074 & 0.56-40.74  & 19-647   & 0.119-1.024 \\
  \cite{Spingola_2020}\tablenotemark{b}                   & no  & yes & 2  & 2.059-3.200 & 25.0-34.0   & 415-542  & 0.461-0.820 \\
  \cite{Dessauges_Zavadsky_2015, Dessauges_Zavadsky_2017}\tablenotemark{b} & no  & no  & 7  & 1.585-3.631 & 0.3-7.4     & 10-81    & 0.076-1.254 \\
  \cite{Oteo2018}                                         & yes & yes & 4  & 4.000-4.000 & 10.8-26.2   & 200-1224 & 0.214-0.539 \\
  \cite{Carilli_2010}\tablenotemark{b}                                    & yes & yes & 1  & 4.050       & 16.0        & 3190     & 0.050 \\
  \cite{Tan_2013}                                         & no  & no  & 2  & 3.216-4.058 & 16.0-18.0   & 181-330  & 0.485-0.996 \\
  \cite{Miller2018}\tablenotemark{b}                                       & yes & yes & 14 & 4.300-4.300 & 1.0-12.0    & 68-1304  & 0.063-0.254 \\
  \cite{Riechers_2010a}                                   & yes & yes & 1  & 5.298       & 5.3         & 1914     & 0.028 \\
  \enddata
  \tablenotetext{a}{Star formation rates for these studies are calculated from the IR luminosity using the \cite{Kennicutt_1998} relation.}
  \tablenotetext{b}{Source galaxies are gravitationally lensed.}
  \tablecomments{All galaxies are CO-detected except for \cite{Zavala2019}. All star formation rates have been scaled to the \cite{Kroupa_2001} IMF following the factors given in \cite{Madau_2014}.}
\end{deluxetable*}

\,
\newpage
\,
\newpage

\Needspace{4\baselineskip}
\section{Moment zero maps}\label{sec:appendix-m0}

Below is an example of a moment zero map. The full figure set containing moment zero maps for all galaxies (22 images) is available in the online version of the paper.

\begin{figure}[!htbp]
  \centering
  \includegraphics[width=1\linewidth]{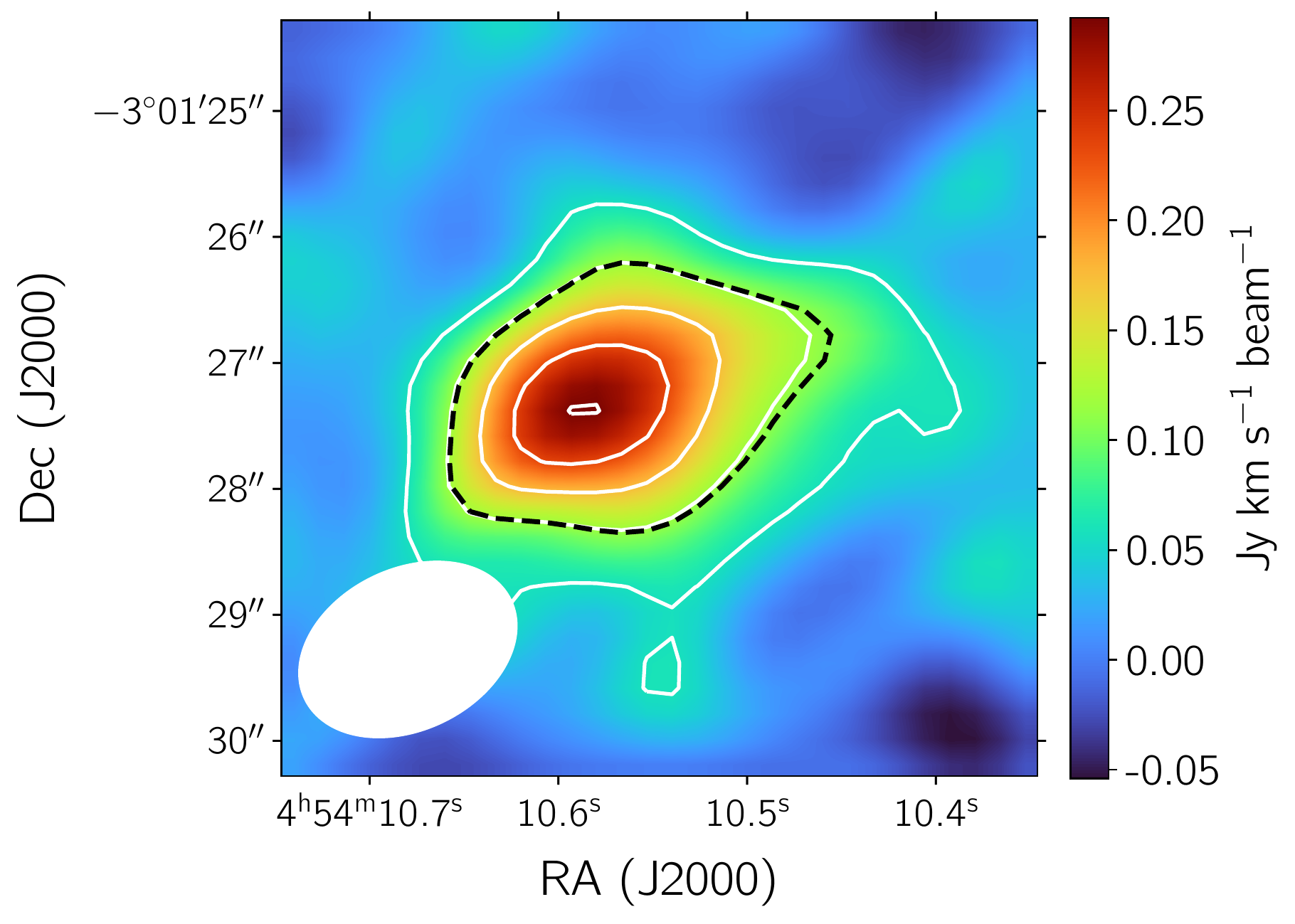}
  \caption{Moment zero map of Gal. 8 centered on the \hst\ position. The image is created from the $-200~{\rm km\,s^{-1}}$ to $200~{\rm km\,s^{-1}}$ channels of the non-primary-beam corrected data cubes, and measures 6\arcsec $\times$ 6\arcsec. The solid white contours start from $2\,\sigma$ and go up in $2\,\sigma$ increments, and the dotted contours start from $-2\,\sigma$ and go in $2\,\sigma$ increments, where $\sigma$ is the rms noise. The ellipses in the bottom-left corners are the ALMA beam, which has FWHM major and minor axis sizes of 1.810\arcsec\ and 1.275\arcsec\ respectively, and a position angle of $-64.3~{\rm deg}$. The optimized extraction regions are indicated by the dotted black line. The full figure set with the moment zero maps for all galaxies (22 images) is provided in the online version of the paper.}
  \label{fig:m0_figset}
\end{figure}

\,
\newpage

\Needspace{4\baselineskip}
\section{Integrated spectra}\label{sec:appendix-spec}

Below is an example of an integrated spectrum. The full figure set containing spectra for all galaxies (22 images) is available in the online version of the paper.

\begin{figure}[!htbp]
  \centering
  \includegraphics[width=1\linewidth]{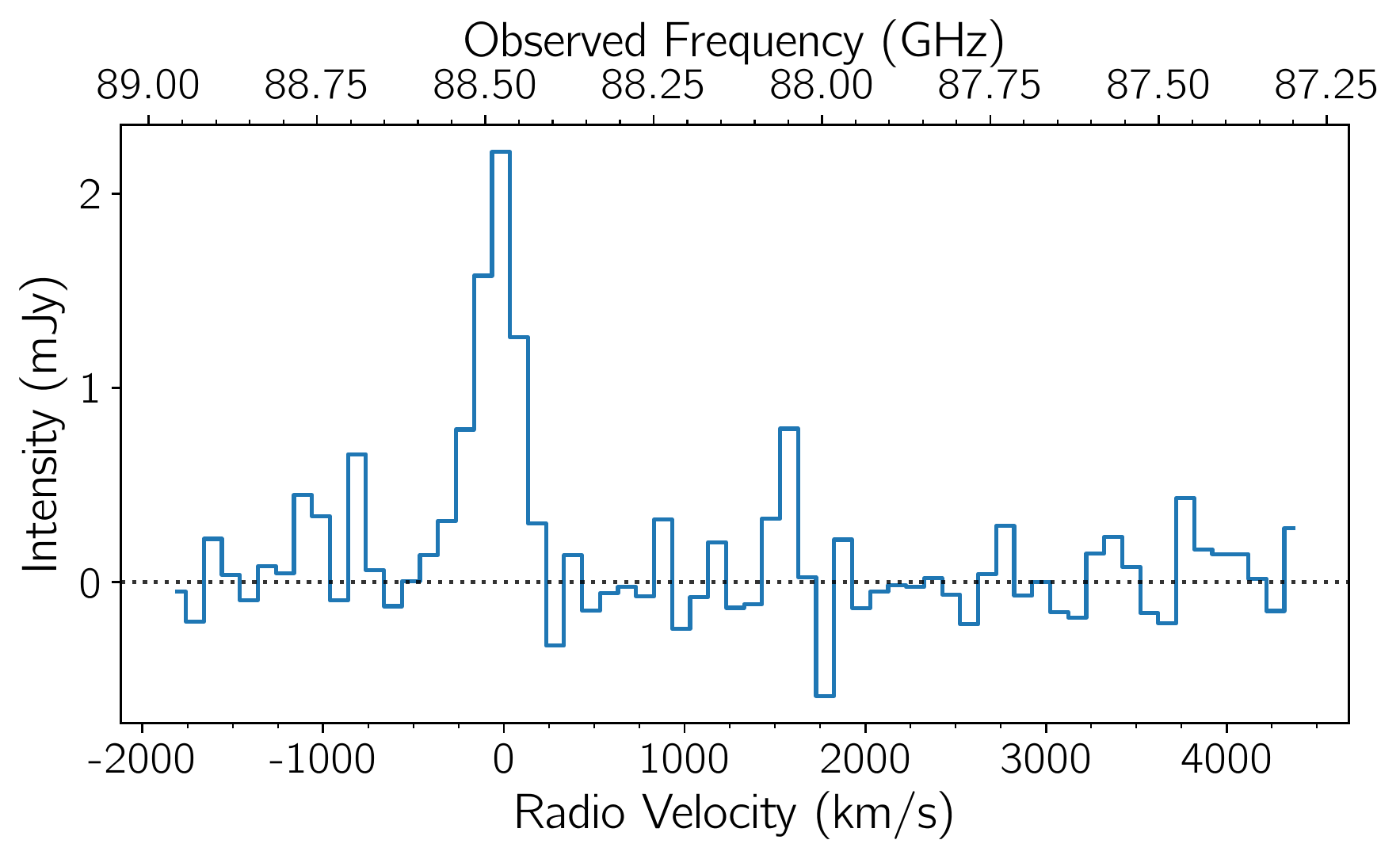}
  \caption{Integrated spectrum of Gal. 8, created by spatially integrating the primary-beam corrected cube in all channels in the spectral window. The integration region is determined by the optimal extraction region (see Section \ref{sec:results-specextraction} for details).}
  \label{fig:spec_figset}
\end{figure}

\,
\newpage

\Needspace{4\baselineskip}
\section{Moment one maps}\label{sec:appendix-m1}

Below is an example of a moment one map. The full figure set containing moment one maps for the detected galaxies (7 images) is available in the online version of the paper.

\begin{figure}[!htbp]
  \centering
  \includegraphics[width=1\linewidth]{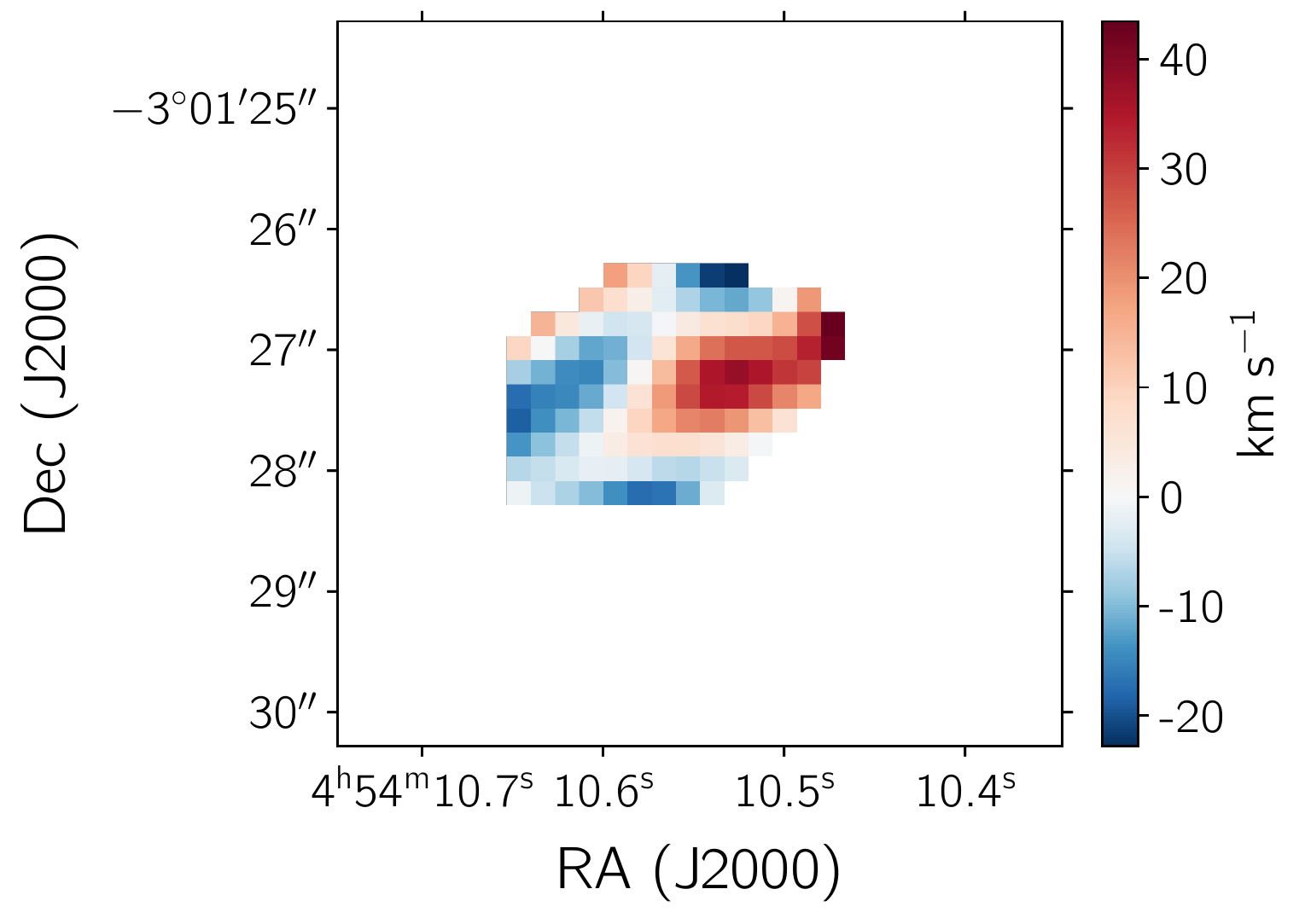}
  \caption{Moment one map of Gal. 8, as created from a data cube masked to only retain the optimal extraction region (see Section \ref{sec:results-specextraction} for details) in the $-200~{\rm km\,s^{-1}}$ to $200~{\rm km\,s^{-1}}$ channels of the non-primary-beam corrected cubes. The velocities are given relative to the systematic velocity of the source.}
  \label{fig:m1_figset}
\end{figure}

\,
\newpage

\Needspace{4\baselineskip}
\section{Continuum images}\label{sec:appendix-cont}

Below is an example of a continuum image. The full figure set containing continuum images for all galaxies (22 images) is available in the online version of the paper.

\begin{figure}[!htbp]
  \centering
  \includegraphics[width=1\linewidth]{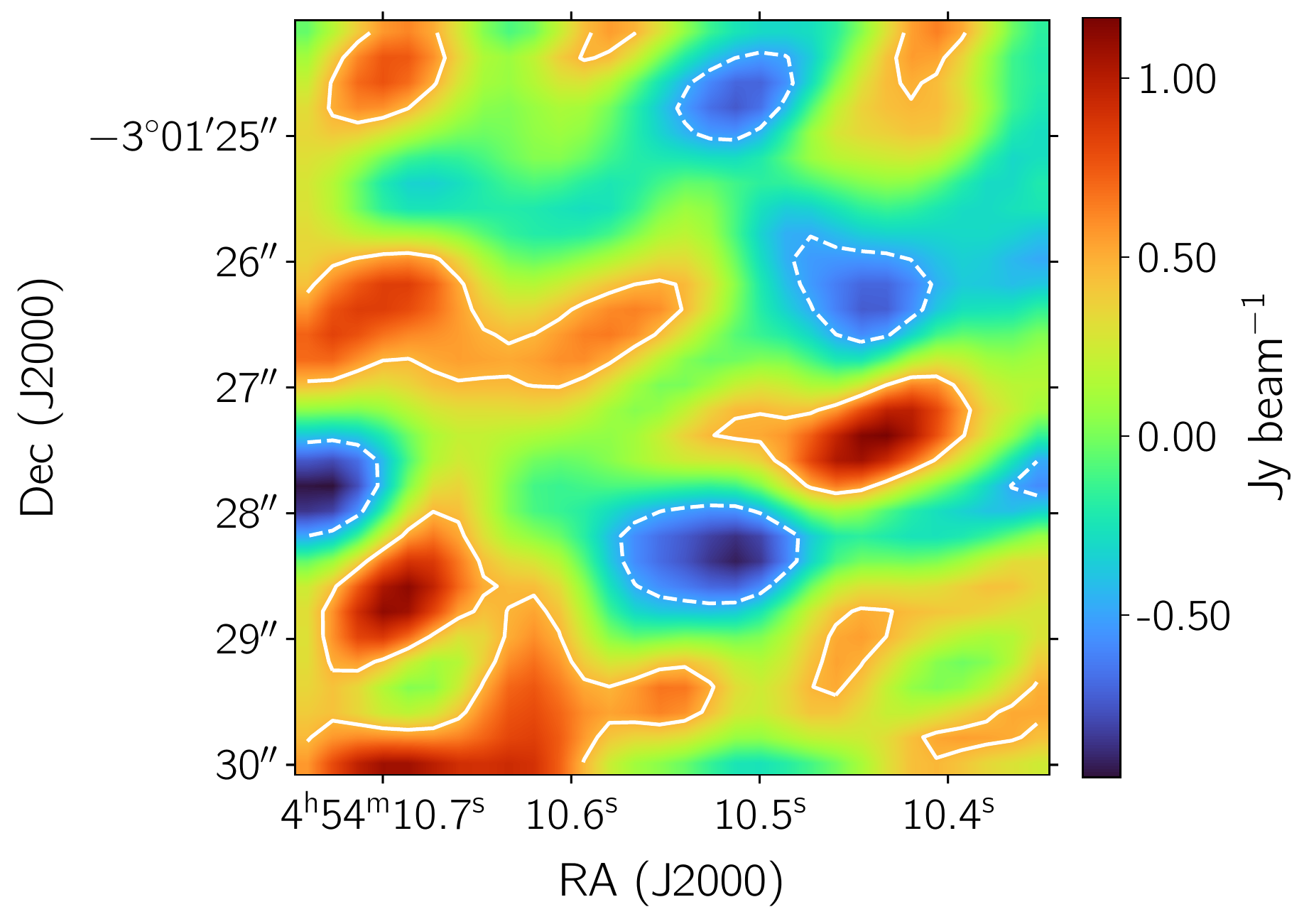}
  \caption{Continuum image of Gal. 8, centered on the \hst\ position and measuring 6\arcsec $\times$\,6\arcsec. The continuum image is created with the three spectral windows where no spectral line emission is expected. The contours indicate the rms noise, with the solid contours starting at $1\sigma$ and increasing in $2\sigma$ increments, and the dashed contours starting at $-1\sigma$ and decreasing in $2\sigma$ increments.}
  \label{fig:cont_figset}
\end{figure}

\end{document}